\documentclass{appolb}
\usepackage{epsfig}
\usepackage{amssymb} 
\def\lsim{\mathrel{\rlap{\lower4pt\hbox{\hskip1pt$\sim$}}
    \raise1pt\hbox{$<$}}}                
\def\gsim{\mathrel{\rlap{\lower4pt\hbox{\hskip1pt$\sim$}}
    \raise1pt\hbox{$>$}}}                
 
\newcommand{\dst}{$D^{*+}$}
\newcommand{\dc}{$D^+$}
\newcommand{\dn}{$D^{\circ}$}
\newcommand{\ds}{$D^+_s$}

\begin{document}
\title{Meson Production and Spectroscopy at HERA
\thanks
{Presented at the MESON2002 Conference 24-28.5.2002, Krak\'{o}w, Poland} 
}
\author{Jan Olsson
\address{DESY, Notkestra\ss e 85, 22607 Hamburg, BRD \\
E-mail: jan.olsson@desy.de \\ \hspace*{1cm} 
\\
On behalf of the H1 and ZEUS collaborations}
}
\maketitle
\begin{abstract}
Selected recent results from the H1 and ZEUS experiments 
are reviewed, illustrating some of the many facets of ``meson physics'' 
at the HERA {\it ep} collider. 
The results cover exclusive
elastic and proton-dissociative diffractive vector meson production 
and comparisons  with recent theoretical calculations
show that perturbative QCD models are successful in
describing these processes when at least one of the involved scales 
have large values. Furthermore 
a search for odderon induced exclusive photoproduction of pseudoscalar and 
tensor mesons is described; upper limits for the cross sections are below 
recent theoretical predictions. Finally  the status of open charm meson 
spectroscopy in inclusive final states is reported.
\end{abstract}
  
\section{Introduction}

In the very successful first running period, which ended in 2000, the 
$ep$ collider HERA at DESY yielded more than 100 pb$^{-1}$ of integrated 
luminosity for each of the two experiments H1 and ZEUS.
These data provide high statistics samples 
in many interesting areas of physics. In the present report,
recent results in three such areas, all related to ``meson''
physics, are described. More precisely, the topics of 
Exclusive Vector Meson Production, 
Production of Pseudoscalar and Tensor Mesons via Odderon Exchange, 
and finally Charm Spectroscopy, are addressed.

\section{Exclusive Vector Meson Production}

The diffractive, exclusive vector meson production
process,
\begin{equation} 
ep \rightarrow\ e V Y,\ \
 {\rm with}\ V = \ \  \rho^{\circ}, \omega,
\phi, J/\Psi, \Psi^{\prime}, \Upsilon,
\end{equation} 
and with $Y$ being either a proton (elastic scattering) or a low mass 
hadronic system (proton dissociative scattering), has been extensively 
studied at HERA. Detailed reviews are given in 
\cite{Crittenden,AbramowiczCaldwell}. 
The renewed interest in this seemingly simple reaction stems from the 
fact that the HERA colliding beam experiments greatly extend the accessible
range both of the center of mass energy and of the physics 
scales which are involved
in the process. Thus detailed studies are possible both in the soft, 
low energy regime already explored by the fixed target experiments, and in 
the new regime, where scales reach values large enough for perturbative QCD
(pQCD) to be applied, \ie  where $\Lambda_{QCD}$ is small in comparison.
\par\noindent
Reaction $(1)$ is shown schematically in Fig. 1a. The virtual
photon, emitted by the scattered electron, scatters diffractively off the
proton to form a vector meson. The following quantities will 
be used in the discussion:    
\vspace*{0.2cm}
\par\noindent
\begin{minipage}[t]{12.6cm}
\begin{minipage}[t]{5cm}
\epsfig{file=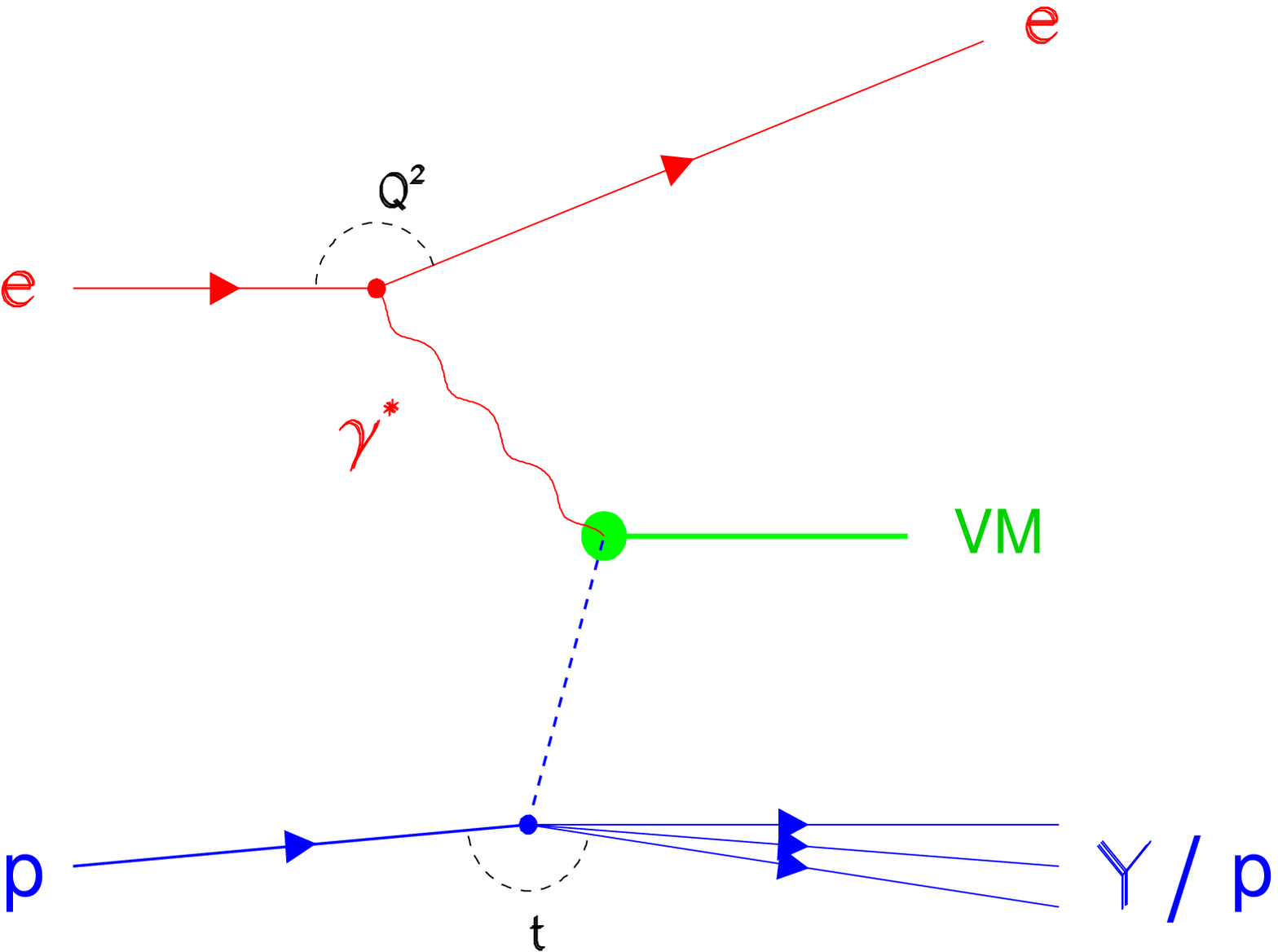,width=5cm}
\vspace*{-0.1cm}
\par\noindent
\epsfig{file=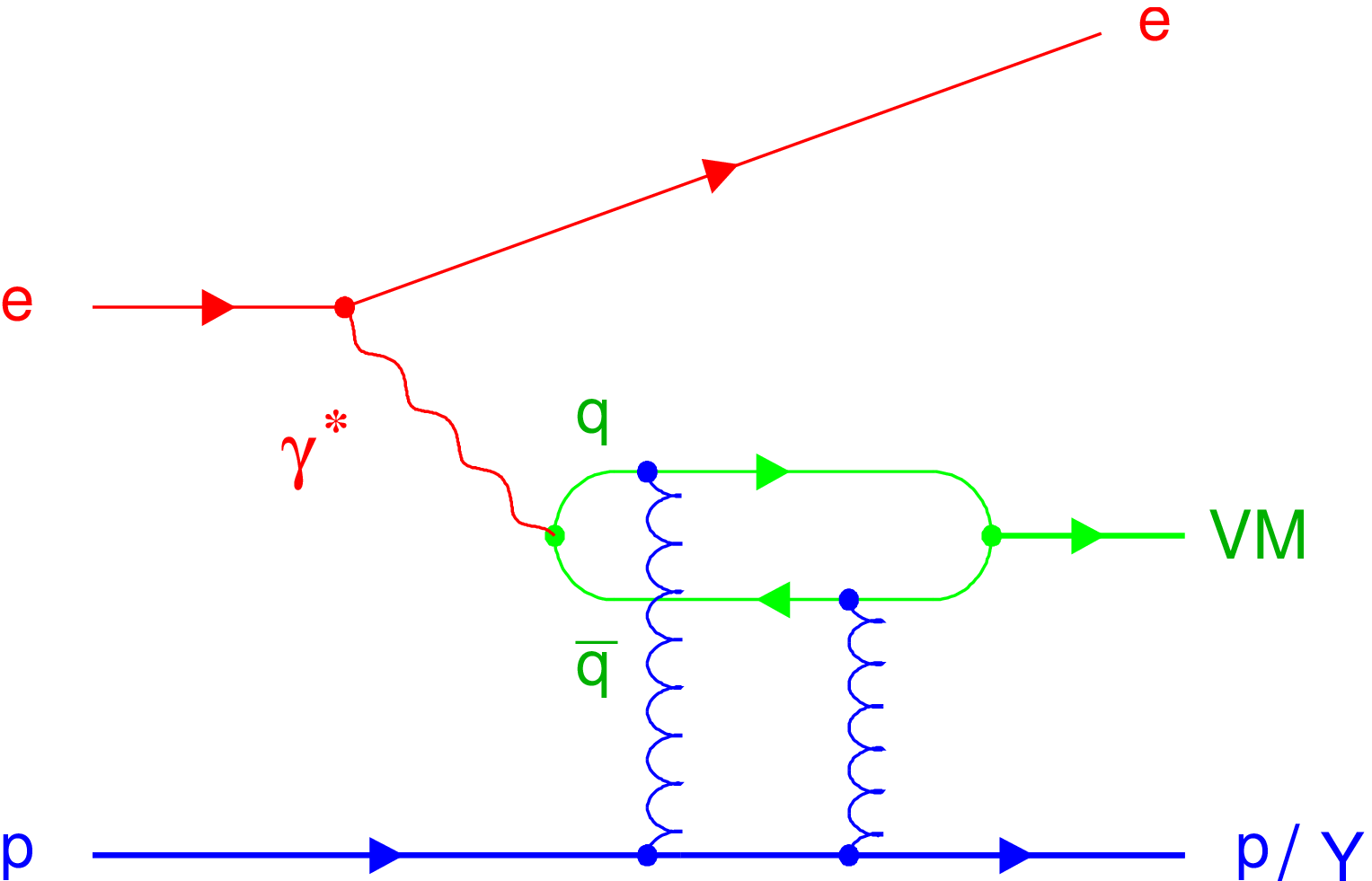,width=5cm}
\par\noindent
Figure 1: Exclusive vector meson production: a) schematic diagram, b) pQCD
approach.
\end{minipage}
\vspace*{-2.6cm}
\par\noindent
\hspace*{0.3cm} b)
\vspace*{-4.6cm}
\par\noindent
\hspace*{0.3cm} a)
\vspace*{-2.4cm}
\par\noindent
\hspace*{5.5cm}
\begin{minipage}[t]{7cm}
\begin{itemize}
\item
$Q^2$: The photon virtuality, i.e. the absolute value of the squared
4-momentum transfer at the electron vertex (the generic term
`electron' is used
for both $e^+$ and $e^-$). In the results presented here
$Q^2$ ranges from $\sim0$ to
$\sim100\ {\rm GeV}^2$.
\item
$W_{\gamma p}$, or simply $W$: 
the CM energy of the $\gamma^{\ast}p$ system, here in
the range \ \ \ $20\lsim W_{\gamma p}\lsim 290\ {\rm GeV}$. 
\item
$t$: The squared 4-momentum transfer at the proton vertex, with
\newline
$0\lsim |t|\lsim 20\ {\rm GeV}^2$ in the present results.
For small values of $Q^2$, $|t|$
is well approximated by the squared transverse momentum of the
produced vector meson $V$, $t\approx -p_{t,V}^2$.
\end{itemize}
\end{minipage}
\end{minipage}
\vspace*{0.5cm} 
\par\noindent
The classical approach, based on Regge theory and Vector Meson
Dominance (VDM), gives a successful description of process $(1)$ for the
light vector mesons and for low values of $Q^2$ and $|t|$.
Diffractive scattering means exchange of 
the vacuum quantum numbers, and the corresponding Regge trajectory is the
Pomeron.
Predictions in this approach are, among others: a slow rise of the cross 
section with $W$, \ie $\sigma \propto W^{0.2 - 0.3}$, shrinkage of the 
diffractive peak with increasing $W$, \ie $ d \sigma/d t\ \propto\
               e^{b t}\, (W/W_0)^{4 (\alpha_{I\!\! P}(t) - 1)}$ with 
      $ \alpha_{I\!\! P}(t)\ =\ \alpha_{I\!\! P}(0)\ +\
           \alpha_{I\!\! P}^{\prime}\, t$ and
$ b\ =\ b_0\ +\ 4\, \alpha_{I\!\! P}^{\prime}\, ln(W/W_0)$, conservation of
the S-channel helicity (SCHC) and a $Q^2$-dependence 
$\sigma \propto 1/(Q^2 + M_{V}^2)^2$.
\vspace*{0.1cm}
\par\noindent
In the pQCD approach, Fig. 1b, the process $(1)$ is seen as 
a series of steps, well separated by the very different 
timescales involved (factorization): 
\begin{itemize}
\item
The virtual photon fluctuates into a $q\bar{q}$ pair, \ie into a colour
dipole.
\item
The $q\bar{q}$ pair scatters off the proton, in leading order 
under exchange of a pair of gluons in a colour singlet state. 
\item
The scattered $q\bar{q}$ pair forms a bound state, the vector meson.
\end{itemize}
\vspace*{0.1cm} 
\par\noindent
\begin{minipage}[t]{12.6cm}
\begin{minipage}[t]{6cm}
While the $q\bar{q}$ scattering can be described within pQCD, the 
$\gamma\rightarrow q\bar{q}$ and $q\bar{q}\rightarrow V$ processes are
modelled with the respective wave-functions. In this approach,
the cross section for $(1)$ is 
proportional to the square of the gluon density
$g(x,Q^2)$ in the proton\cite{Brodsky94},
\begin{equation}
\sigma \propto \alpha_s^2(Q^2)/Q^6\,\, [x g(x,Q^2)]^2,
\end{equation}
where $x$ is the Bjorken $x$. pQCD predictions for reaction $(1)$ are
partly different from those of the Regge and VDM approach. 
Thus the cross section will rise
steeply with $W$, due to the increasing gluon density in the proton at
small values of $x$ and the relation $xW_{\gamma p}^2\approx Q^2$, valid
for small $x$ at a given $Q^2$. Furthermore, much reduced shrinkage of the 
diffractive peak is expected, and SCHC is violated. The $Q^2$-dependence in
$(2)$ is discussed below.
\par\noindent
The HERA data have been extensively 
used to demonstrate the va-
\end{minipage}
\vspace*{-11.7cm}
\par\noindent
\hspace*{6.4cm}
\begin{minipage}[t]{6.1cm}
\epsfig{file=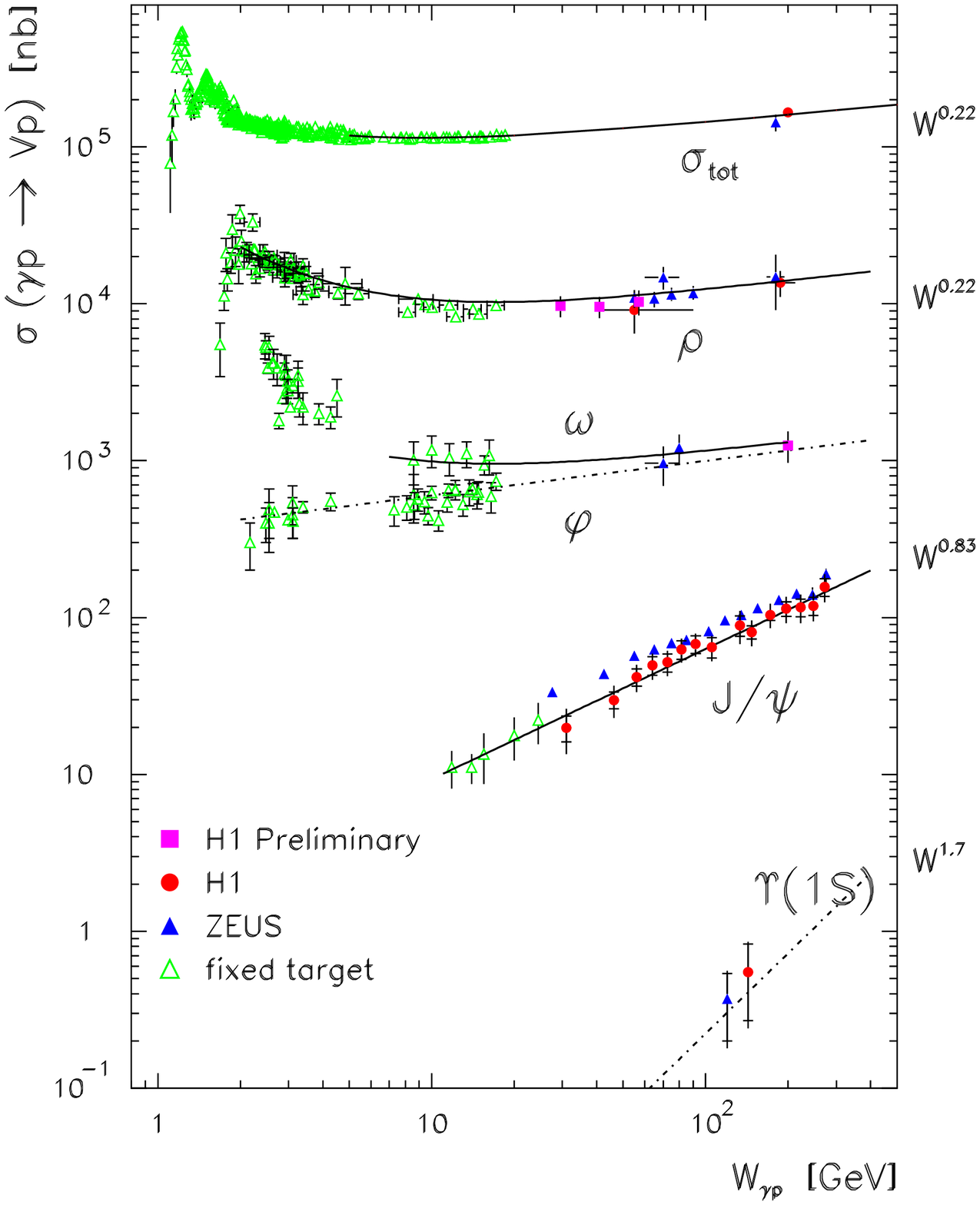,width=6.1cm,
      bbllx=20pt,bblly=60pt,bburx=570pt,bbury=730pt,clip= }
Figure 2: Compilation of cross sections for exclusive photoproduction
of the vector mesons $\rho^{\circ}, \omega, \phi, J/\Psi$ and $\Upsilon$, as 
functions of $W_{\gamma p}$. Full curves represent fits using Regge
parametrizations and single pomeron exchange. Dashed lines
indicate $W_{\gamma p}$-dependences as given on the right hand side.
\end{minipage}
\end{minipage}
\vspace*{0.05cm}
\par\noindent
lidity of the pQCD approach, 
as the following examples illustrate.
\vspace*{0.2cm}
\par\noindent
{\bf \boldmath$W$-dependence of the cross sections: \ \ }
Fig. 2 shows a compilation of cross section measurements
$\sigma_{\gamma p\rightarrow Vp}$ vs. $W$, for
exclusive photoproduction of vector mesons at HERA and at
lower (fixed target) CM energies. Also the total photoproduction cross
section is shown. The slow rise with $W$, as predicted in the Regge approach,
is clearly seen for the light vector mesons $\rho, \omega$ and $\phi$ (and 
also for the total cross section). However, the rise with $W$ is much steeper
for the heavier $J/\Psi$. The observed $W^{\delta}$-dependence can easily 
be related to the increasing proton gluon density at small $x$ values:
using Ryskin's\,\cite{Ryskin93} proposed scale
$\mu^2=(Q^2+M^2_{J/\Psi})/4$ for photoproduction at small $|t|$,
a value $\mu^2=2.4\ {\rm GeV}^2$ is obtained. At this scale, and at
small $x$, the gluon
density rises\,\cite{H1ZEUS9596} as $xg(x)\propto x^{-0.2}$,
thus corresponding to $\sigma \propto W^{0.8}$, which is in good
agreement with the observed $\delta\sim 0.7-0.8$.
This indicates that the
mass of the $c$-quark can be used as a hard scale in pQCD 
calculations.
However, 
theory uncertainties currently limit the 
access to the 
\vspace*{-0.7cm}
\par\noindent
\begin{minipage}[t]{12.6cm}
\begin{minipage}[t]{5cm}
\epsfig{file=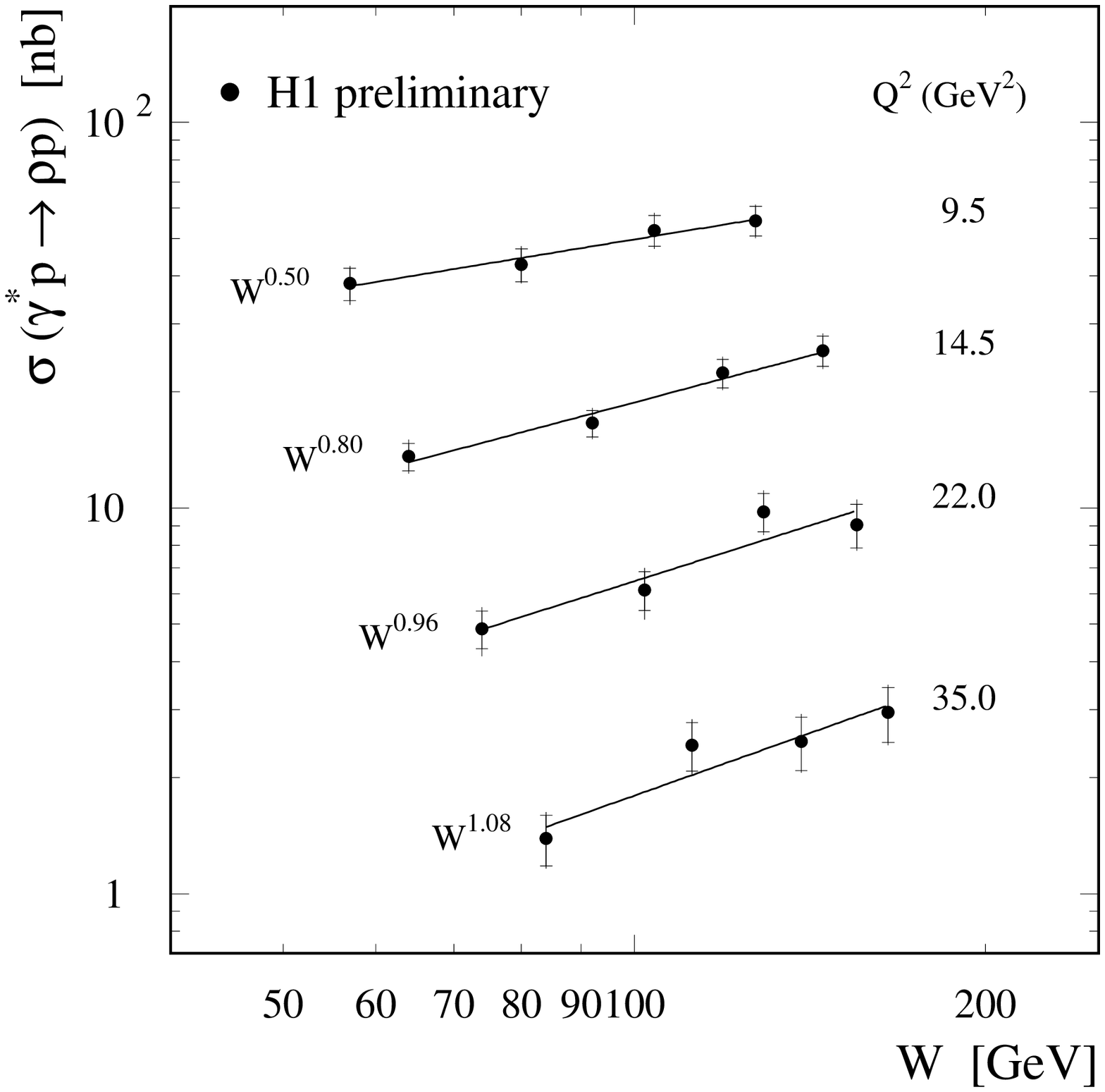,width=5cm}
\par
\epsfig{file=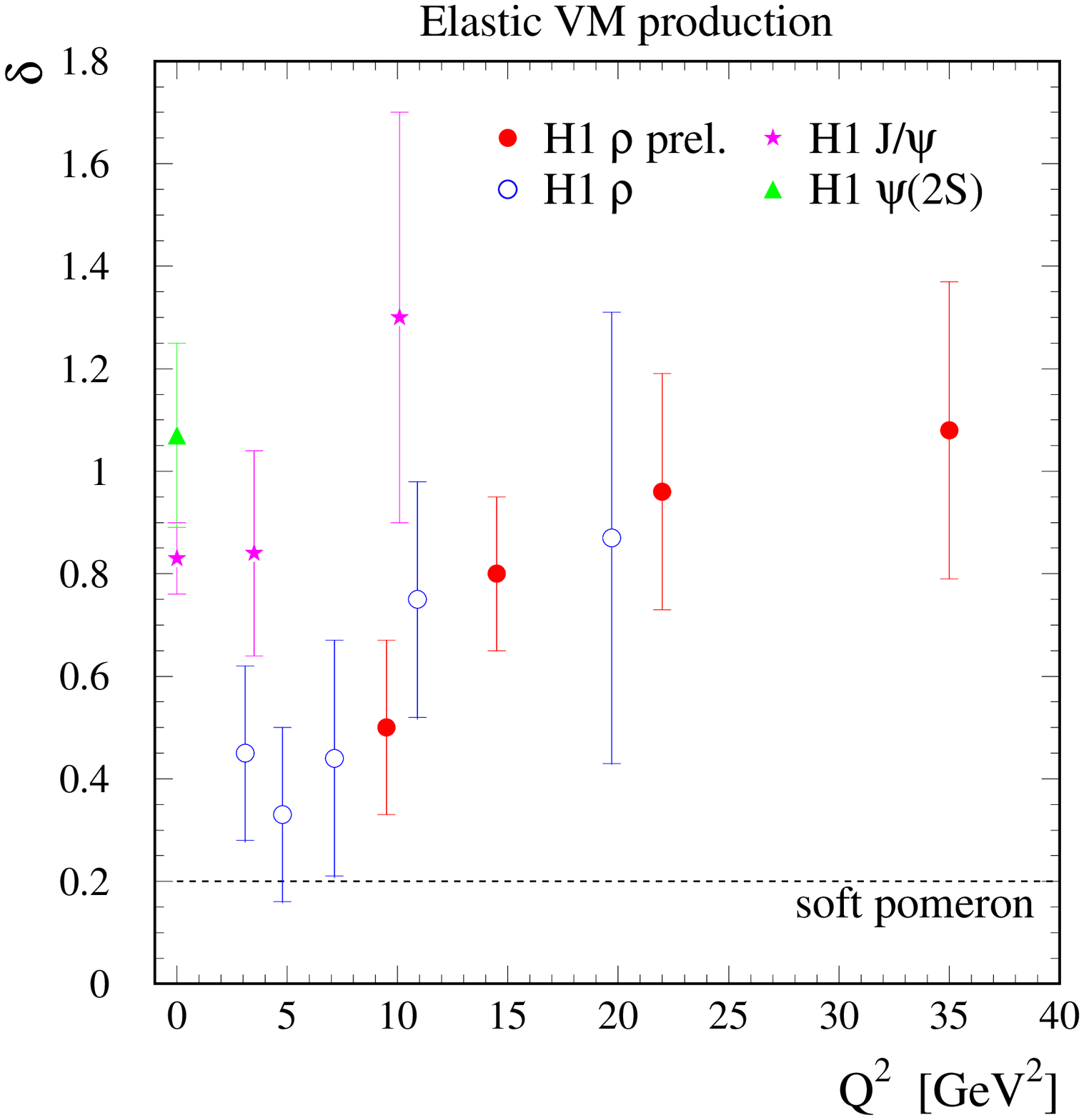,width=4.5cm,
      bbllx=5pt,bblly=5pt,bburx=525pt,bbury=540pt,clip= }
Figure 3: a) Cross section for exclusive $\rho^{\circ}$ production,
as function of $W$ for several values of $Q^2$. Curves show fits
to the form $W^{\delta}$. b) Values of $\delta$ vs. $Q^2$, for $\rho^{\circ}, 
J/\Psi$ and $\Psi^{\prime}$.
 The dotted line shows the ``soft pomeron'' expectation. 
\end{minipage}
\vspace*{-5.4cm}
\par\noindent
\hspace*{3.6cm} b)
\vspace*{-4.8cm}
\par\noindent
\hspace*{3.6cm} a)
\vspace*{-3.8cm}
\par\noindent
\hspace*{5.5cm}
\begin{minipage}[t]{7.1cm}
proton gluon density via precise measurements of the 
$W$-dependence of the cross section for exclusive heavy vector meson 
production.
\vspace*{0.1cm}
\par\noindent
What happens to the $W$-dependence of the cross section for light
vector mesons, if another scale, \eg $Q^2$, is increased? This is 
illustrated in Fig.~3a, where the cross section for  
$\gamma^{*} p \rightarrow  \rho^{\circ}p$ is shown vs. $W$, for increasing 
average values of $Q^2$\cite{H1Newrhodata}. 
The cross section steepens in $W$ with increasing
$Q^2$, and soon deviates significantly 
from the ``soft pomeron''\cite{DL84} expectation,
as seen in Fig.~3b where the power $\delta$ from the $W^{\delta}$ fits
is plotted vs. $Q^2$. At $Q^2>10$ GeV$^2$ the values of $\delta$ for
the $\rho^{\circ}$ cross section are the same as for the $J/\Psi$ 
(or $\Psi^{\prime})$ cross section at $Q^2\sim 0$.
\vspace*{0.2cm}
\par\noindent
The $W$-dependence of the $J/\Psi$ cross section, for increasing  
$Q^2$, is shown in Fig.~4a\cite{H1ZEUSJPsidata}. 
There is no significant change when going 
to higher $Q^2$ values, the already ``hard'' behaviour at $Q^2\sim~0$ does not
become still harder. 
The data are well described by the pQCD calculations\cite{FKS98,MRT99},
using recent proton parton density parametrizations\cite{CTEQ45M}.
\vspace*{0.1cm}
\par\noindent
The $W$-dependence of the $J/\Psi$ cross section also does not change 
when the other 
\end{minipage}
\par\noindent
scale, $|t|$, increases in value. This is seen in
Fig.~4b,  where the cross section 
\end{minipage}
is plotted vs. $W$ for two intervals 
of $|t|$\cite{Browndis2001}. The pQCD model calculation 
in Fig.~4b\cite{Forshaw96} agrees with the data. 
\vspace*{0.1cm}
\par\noindent
{\bf \boldmath$Q^2$-dependence of the cross section: \ \ }
The cross section $(2)$ contains the factor $1/Q^6$, which is much steeper
than the classical, VDM expectation $1/(Q^2 + M_{V}^2)^2$. 
However, the $1/Q^6$ dependence in $(2)$ 
is modified by the $Q^2$-dependence in $\alpha_s$ and in the proton gluon
density. Effectively, a $1/Q^n$ dependence is predicted, with $n \approx 4-5$,
depending on the $Q^2$ range. This is also borne out in the data. Fig.~5a
shows the $Q^2$-dependence of the $\rho^{\circ}$
 cross section with a fit to the 
form $1/(Q^2 + M_{\rho^{\circ}}^2)^n$, 
with $n=2.60$ in the chosen fit interval\cite{H1Newrhodata}. It
is clear that the whole spectrum cannot be fitted with one simple curve of
this form, and
that at low values of $Q^2$ the dependence is less steep. 
\vspace*{-0.2cm}
\par\noindent    
\begin{minipage}[t]{12.6cm}
\hspace*{0.4cm}
\begin{minipage}[t]{4.8cm}
\epsfig{file=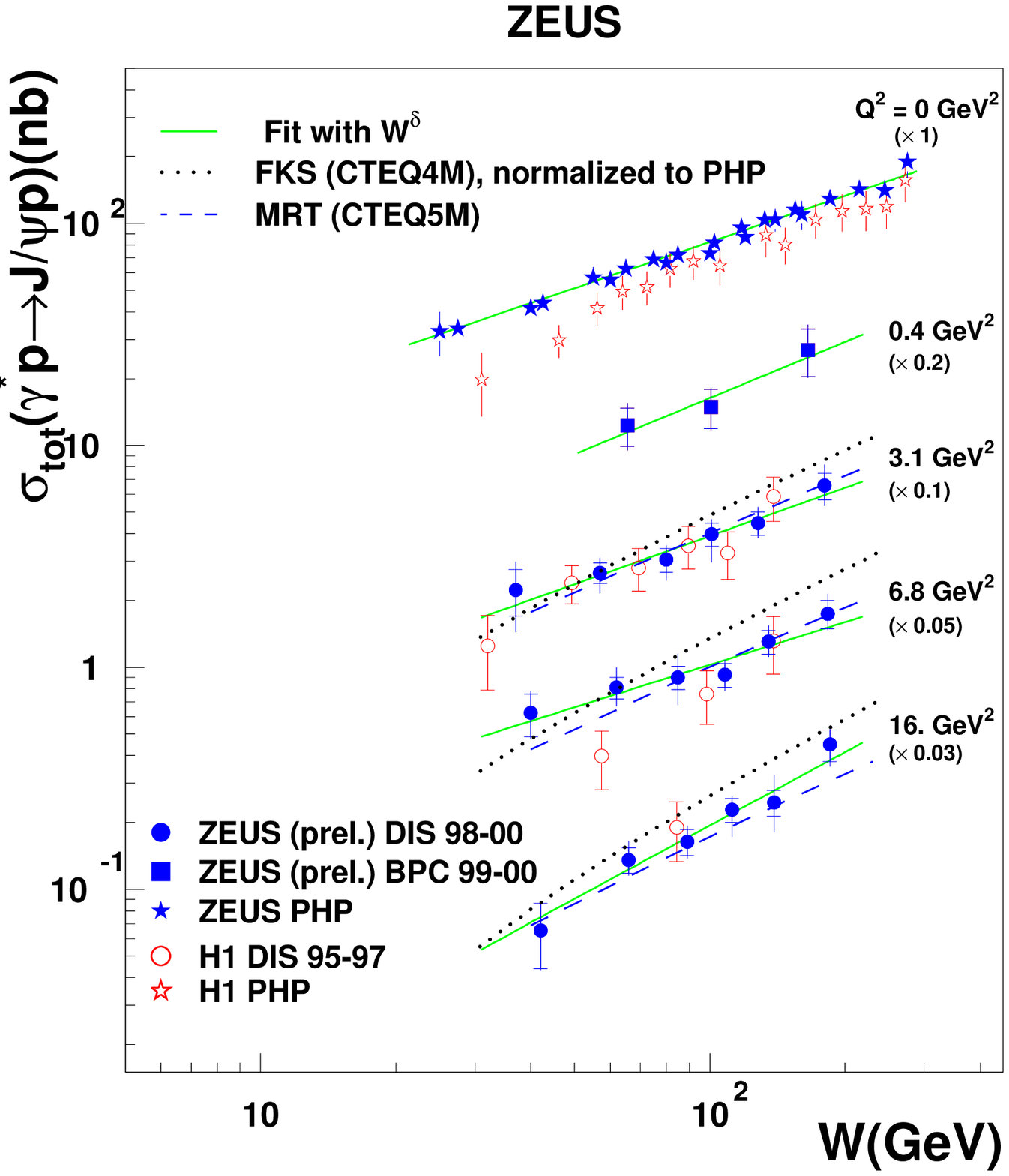,width=4.8cm,
      bbllx=0pt,bblly=10pt,bburx=460pt,bbury=545pt,clip= }
\end{minipage}
\vspace*{-5.3cm}
\par\noindent
\hspace*{6cm}
\begin{minipage}[t]{5.6cm}
\epsfig{file=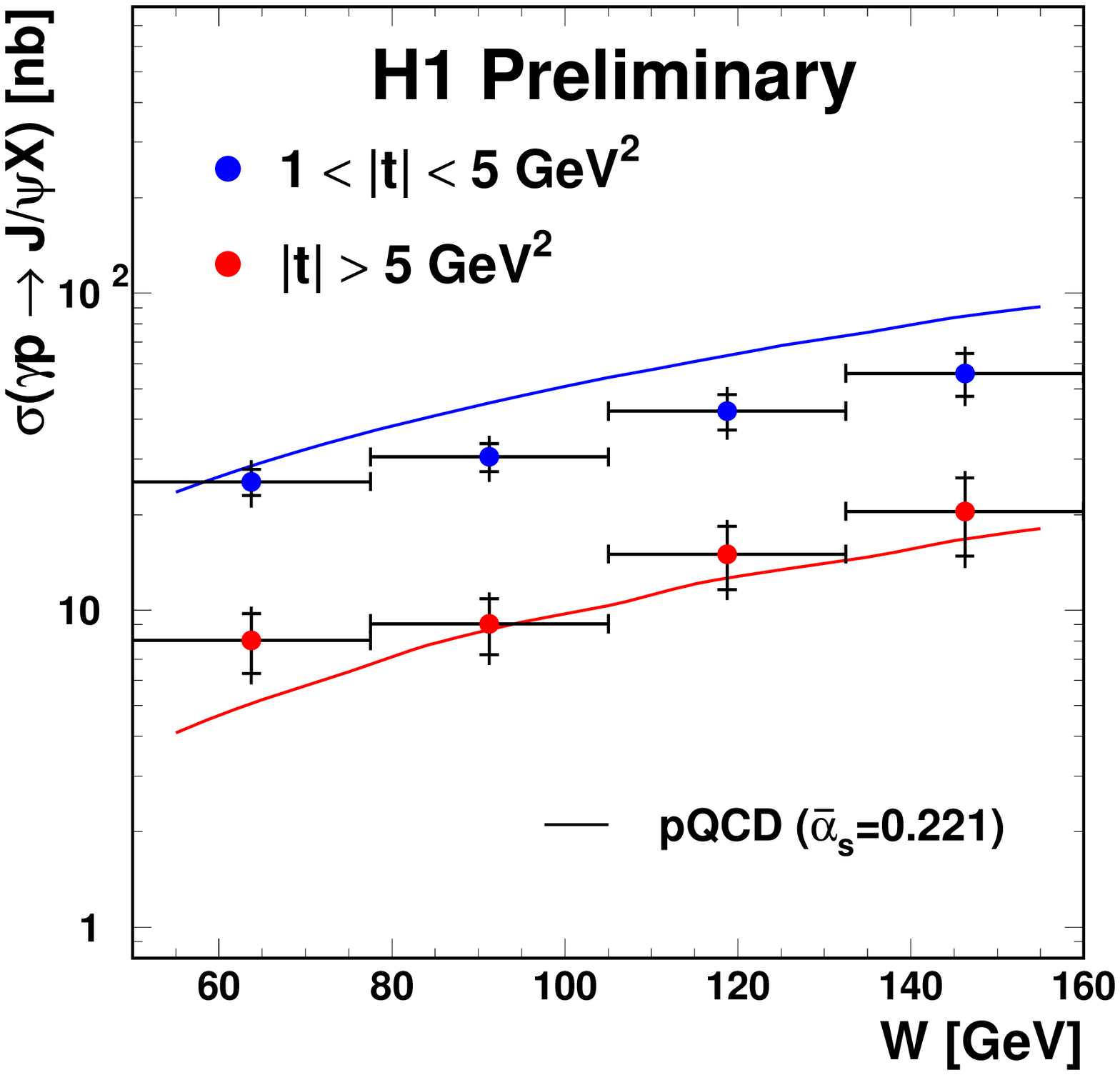,width=5.6cm,
         bbllx=30pt,bblly=160pt,bburx=525pt,bbury=635pt,clip= }
\end{minipage}
\par\noindent
Figure 4: Cross section for exclusive  
$J/\Psi$ production,
as a function of $W$. \newline
a) For several values of $Q^2$. 
Curves show fits to the form $W^{\delta}$ as well as two pQCD 
calculations\cite{FKS98,MRT99}.
b) Photoproduction, for two intervals of $|t|$. The curves show a pQCD
calculation\cite{Forshaw96}.
\vspace*{-3.4cm}
\par\noindent
\hspace*{4.5cm} a) \hspace*{2cm} b)
\end{minipage}
\vspace*{1.9cm}
\par\noindent
\begin{minipage}[t]{12.6cm}
\hspace*{0.3cm}
\begin{minipage}[t]{5cm}
\epsfig{file=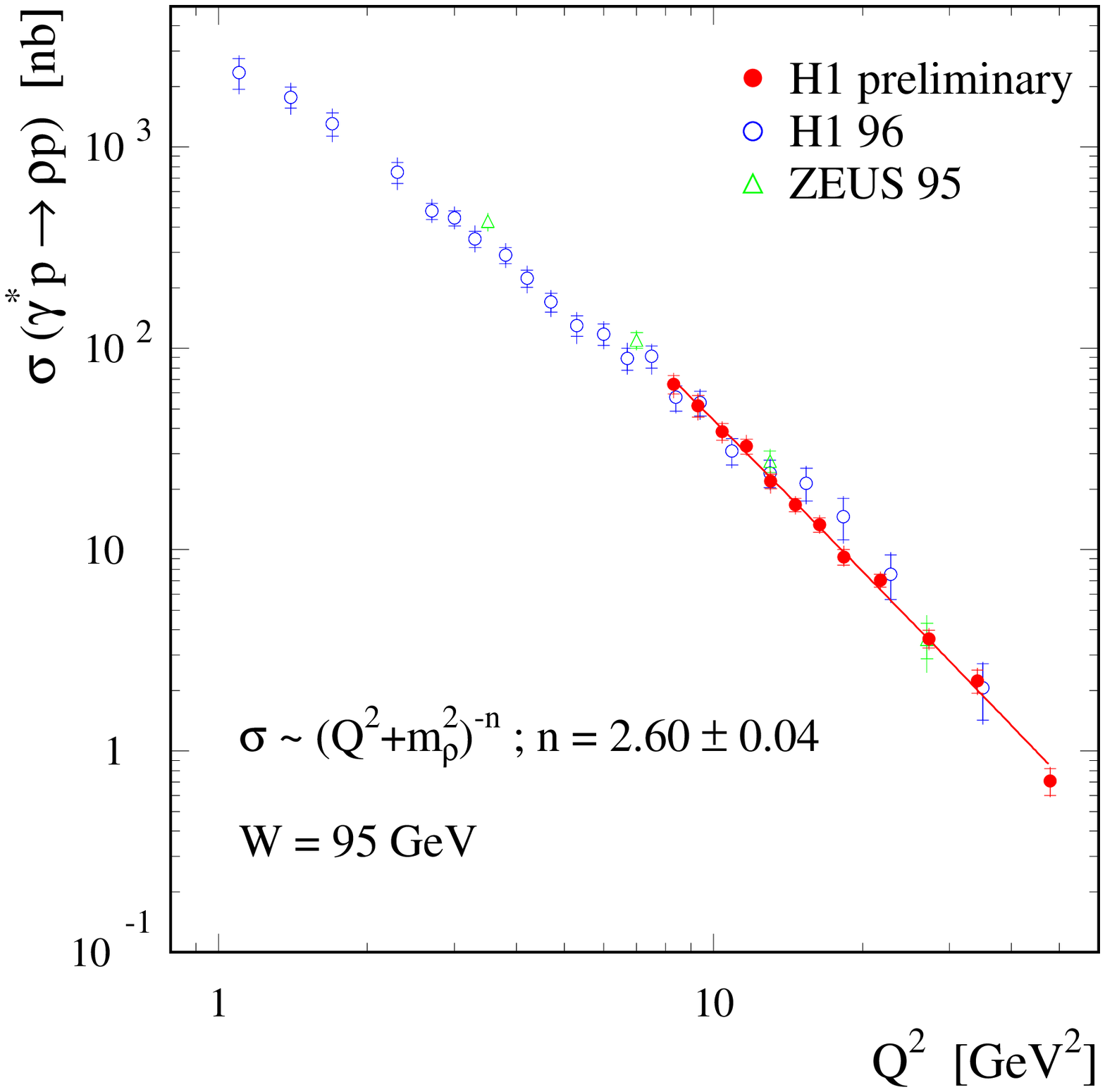,width=5cm,
      bbllx=0pt,bblly=0pt,bburx=515pt,bbury=515pt,clip=} 
\end{minipage}
\vspace*{-5.3cm}
\par\noindent
\hspace*{6.5cm}
\begin{minipage}[t]{5cm}
\epsfig{file=olsson5b.eps,width=5cm,
      bbllx=5pt,bblly=10pt,bburx=465pt,bbury=485pt,clip=}
\end{minipage}
\par\noindent
Figure 5: $Q^2$ dependence of the cross section for exclusive $\rho^{\circ}$ 
production. In a) the total $\rho^{\circ}$ cross section is shown, 
in b)  $\sigma_L$ and $\sigma_T$ are shown separately. 
\vspace*{-5cm}
\par\noindent
\hspace*{4.5cm} a) \hspace*{5.2cm} b)
\end{minipage}
\newpage
\par\noindent
Expression $(2)$ is only valid for the longitudinal part of the cross
section, $\sigma_L$. In \cite{Brodsky94} also the transverse part, $\sigma_T$,
is calculated and predicted to
have a still steeper $Q^2$-dependence, by another factor
$1/Q^2$. Thus $\sigma_L$ is expected to dominate at larger $Q^2$-values.
Data confirm this prediction; in Fig.~5b 
the $Q^2$-dependences of $\sigma_L$ and $\sigma_T$ are shown 
separately\cite{ZEUSLT}.
Indeed, the $Q^2$-dependence of $\sigma_L$
is even harder than the VDM expectation ($n=2$). 
\par\noindent    
The steepness of the $Q^2$-dependence of the cross
section is also modified by \eg the Fermi motion of the
quarks\,\cite{Brodsky94,FKS96} and the suggestion has been made 
that, beyond the pQCD tests,
precise measurements of the $Q^2$-dependence
of the cross section for elastic vector meson electroproduction can reveal
information also about the wave functions of the vector mesons. 
\vspace*{0.2cm}
\par\noindent
\begin{minipage}[t]{12.6cm}
\begin{minipage}[t]{5cm}
\epsfig{file=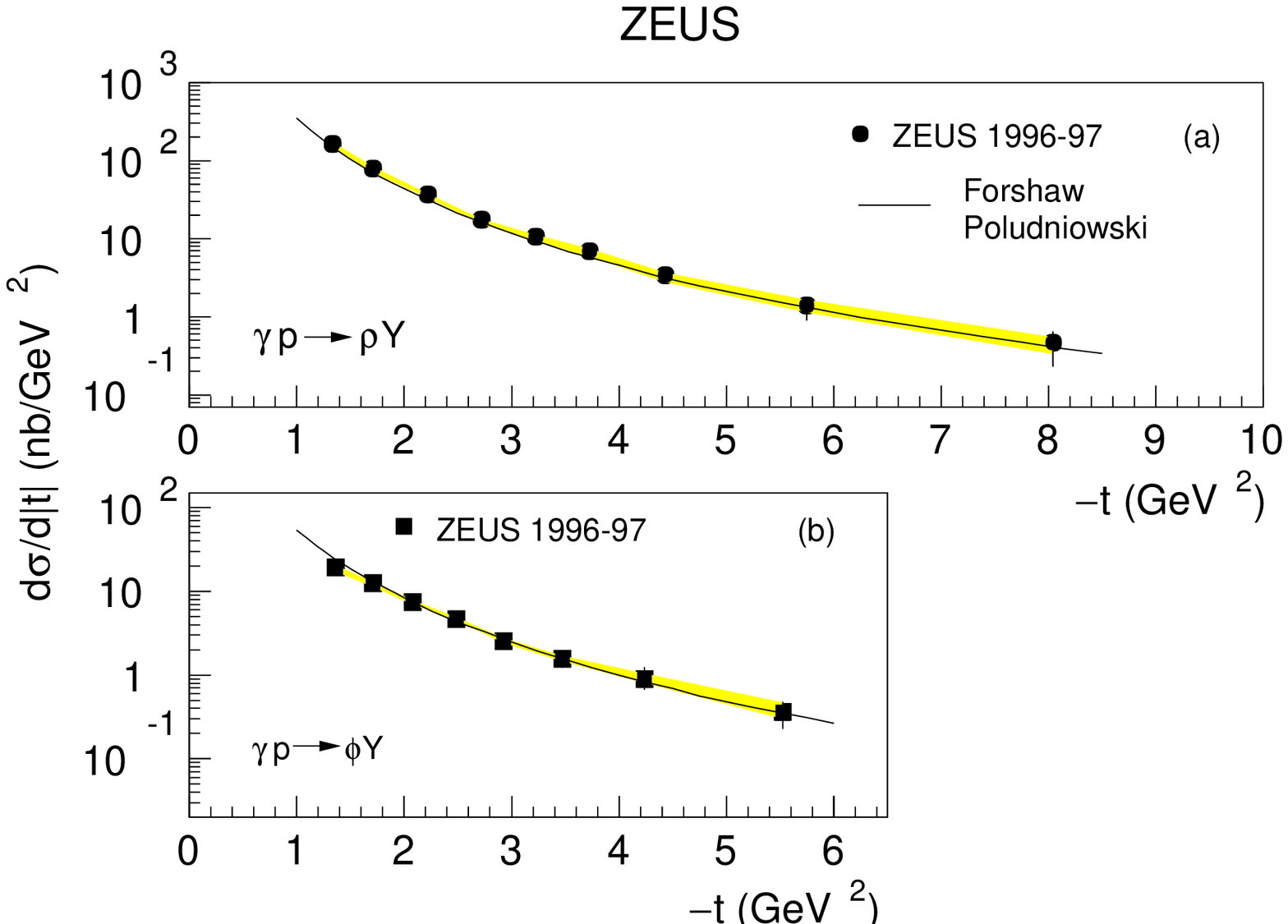,width=5cm}
\vspace*{0.2cm}
\par\noindent
\epsfig{file=olsson6b.eps,width=4.5cm,
      bbllx=5pt,bblly=10pt,bburx=465pt,bbury=485pt,clip=}
Figure 6: $t$-dependence of exclusive $\rho^{\circ}$ and $\phi$ (a-b) and 
$J/\Psi$ (c)  photoproduction cross sections.   Curves
show the pQCD models \cite{Forshawnew} (a-b) and \cite{Forshaw96} (c).
\end{minipage}
\vspace*{-6.1cm}
\par\noindent
\hspace*{4cm} c)
\vspace*{-6cm}
\par\noindent
\hspace*{5.5cm}
\begin{minipage}[t]{7cm}
{\bf \boldmath$t$-dependence of the cross section: \ \ }
New measurements\cite{Browndis2001,newhightdata} from ZEUS and H1 
of the $t$-dependence at large $|t|$
of the exclusive vector meson production cross section
are shown in Fig.~6 for $\rho^{\circ}, \phi$ and $J/\Psi$.
pQCD calculations for the light vector mesons\cite{Ivanov2000} predict
a dependence $d \sigma/d t \propto |t|^{-n}$, $n$ taking values from $\sim$3.8
to $\sim$4.8. Data however exhibit a flatter behaviour. The curves in 
Figs.6a and 6b show the pQCD calculations in \cite{Forshawnew}, which is an 
extension of the BFKL\cite{BFKL} approach taken in \cite{Forshaw95,Forshaw96}.
Both $\rho^{\circ}$ and $\phi$ data are well described. 
This is also true for the heavy quark calculation of \cite{Forshaw96}, 
shown for the $J/\Psi$ data in Fig.6c.  
\vspace*{0.1cm}
\par\noindent
{\bf Helicity studies: \ \ }
The angular distributions involved in the
production and decay of the vector mesons $V$ in the reaction
$\gamma^{\ast}p\rightarrow V\,p$ provide information about the
polarization states of the photon and $V$. Studies
of these angular distributions
are particularly interesting at large values of $Q^2$ and $|t|$, since
pQCD models (\eg \cite{RCKNZ,IvanovKirschner}) 
make predictions about the polarization in these regimes.
Thus dominance is expected 
of longitudinally polarized vec- 
\end{minipage}
\end{minipage}
\par\noindent
tor mesons produced by
longitudinally polarized photons and violation of SCHC is also predicted. 
\par\noindent
Three angles are defined in the helicity system, 
the commonly used reference frame:
\begin{itemize}
\item
$\Phi$:\ \ in the hadronic ($\gamma p$) CM the azimuthal angle between the
electron scattering plane and the
plane containing $V$ and the scattered proton,
\item
\vspace*{-0.1cm}
$\theta^*$:\ \ the decay angle of the 2-body decay of $V$, defined
 by the positive decay particle in the rest system of $V$, with the
quantization axis taken as the direction of $V$ in the
$\gamma p$ CM,
\item
\vspace*{-0.1cm}
$\varphi$:\ \ the azimuthal angle of the positive decay particle in the
rest system of $V$, \ie the angle between the $V$ decay and the $V$ 
production planes.  
\end{itemize} 
\par\noindent
The normalized angular distribution $W(\cos\theta^*,\varphi,\Phi)$ can be
written\,\cite{SWolf}
as a function of 15 quantities $r^{\alpha}_{ik}, r^{04}_{ik}$,
which are linear combinations of the spin density matrix elements (SDMEs). 
The subscripts $i$ and $k$ take the values of
the possible helicity states $-1, 0, 1$. The superscripts 0 and 4 refer
respectively to
unpolarized transverse photons and longitudinally polarized photons,
superscripts $\alpha=1$ and $2$ correspond to
linearly polarized transverse photons, and superscripts $\alpha=5$
and $6$ represent
the interference of transverse and longitudinal amplitudes. 
\par\noindent
In the case of SCHC and natural parity
exchange (NPE) the 15 independent SDMEs
are constrained, and only 5 elements are non-zero.
\vspace*{0.4cm}
\par\noindent
\begin{minipage}[t]{12.6cm}
\begin{minipage}[t]{5cm}
\epsfig{file=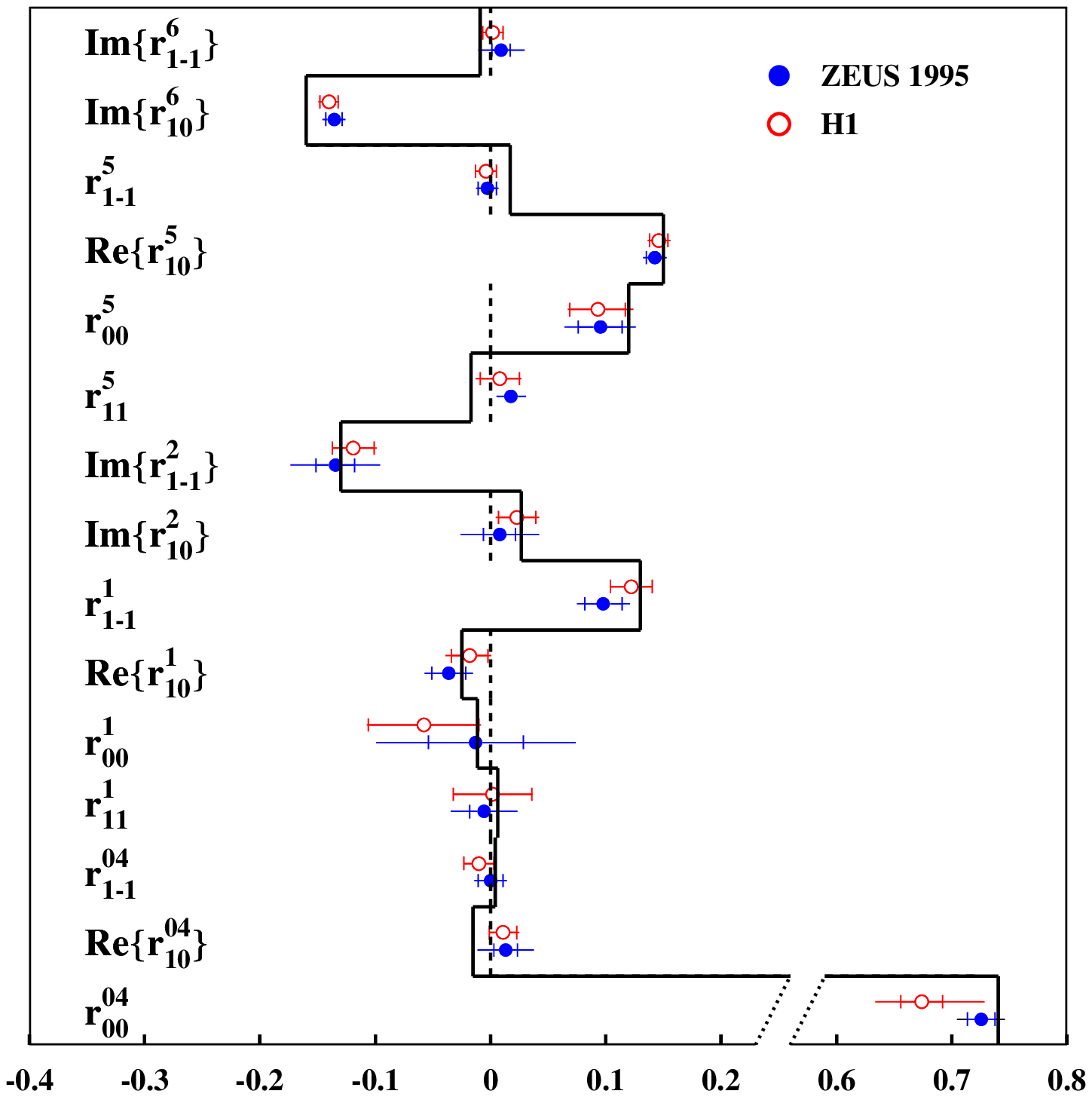,width=5cm,
   bbllx=40pt,bblly=40pt,bburx=460pt,bbury=460pt,clip= }
Figure 7: 15 SDMEs from exclusive $\rho^{\circ}$ 
electroproduction data. 
Full and dotted lines: pQCD calculation\cite{IvanovKirschner} 
and SCHC expectation. 
\end{minipage}
\vspace*{-7.5cm}
\par\noindent
\hspace*{5.5cm}
\begin{minipage}[t]{7cm}
H1 and ZEUS have performed large statistics helicity
analyses of elastic $\rho^{\circ}$\,\cite{H1elprodrho}
and $\phi$\,\cite{ZEUSelprodphi} electroproduction.  
The result for the $\rho^{\circ}$ analysis
is shown for the two experiments in Fig.~7, where all 15 SDMEs
have been determined. 
The prediction from SCHC and NPE is 
shown as dotted lines at zero.
The two experiments (which are in very good agreement with each other)
deviate significantly from this prediction in one element, namely  
$r^5_{00}$ which is clearly non-zero.
This deviation is predicted by the pQCD model of 
Ivanov and Kirschner\,\cite{IvanovKirschner}, which is everywhere 
in excellent agreement with the data.
The element $r^5_{00}$ is approximately proportional to the amplitude
$T_{01}$ for a transverse photon to produce a longitudinal vector meson.
\end{minipage}
\end{minipage}
\vspace*{0.2cm}
\par\noindent
Recently, H1 and ZEUS have extended the helicity studies of 
$\rho^{\circ}$ production to larger values of
$Q^2$ and $|t|$\cite{H1Newrhodata,newhightdata,H1rhohight}. 
The H1 analysis studies the single angular distributions
\begin{equation}
W(\Phi) \propto\ 1 - \epsilon \cos 2\Phi (2r^1_{11} + r^1_{00})
+ \sqrt{2\epsilon (1+\epsilon )}\cos \Phi (2r^5_{11} + r^5_{00})
\end{equation}
and
\begin{equation}
 W(\cos \theta^*) \ \sim \  1 -  r^{04}_{00}\  +
 \ (3\, r^{04}_{00} - 1) \cos^2 \theta^*
\end{equation}
in each of which the two remaining angles have been integrated over. 
The polarization parameter \, $\epsilon$ \, has the value 
$\approx 0.99$ in these analyses.
\vspace*{0.3cm}
\par\noindent
\begin{minipage}[t]{12.6cm}
\begin{minipage}[t]{7cm}
\begin{minipage}[t]{7cm}
\hspace*{-0.2cm}
\epsfig{file=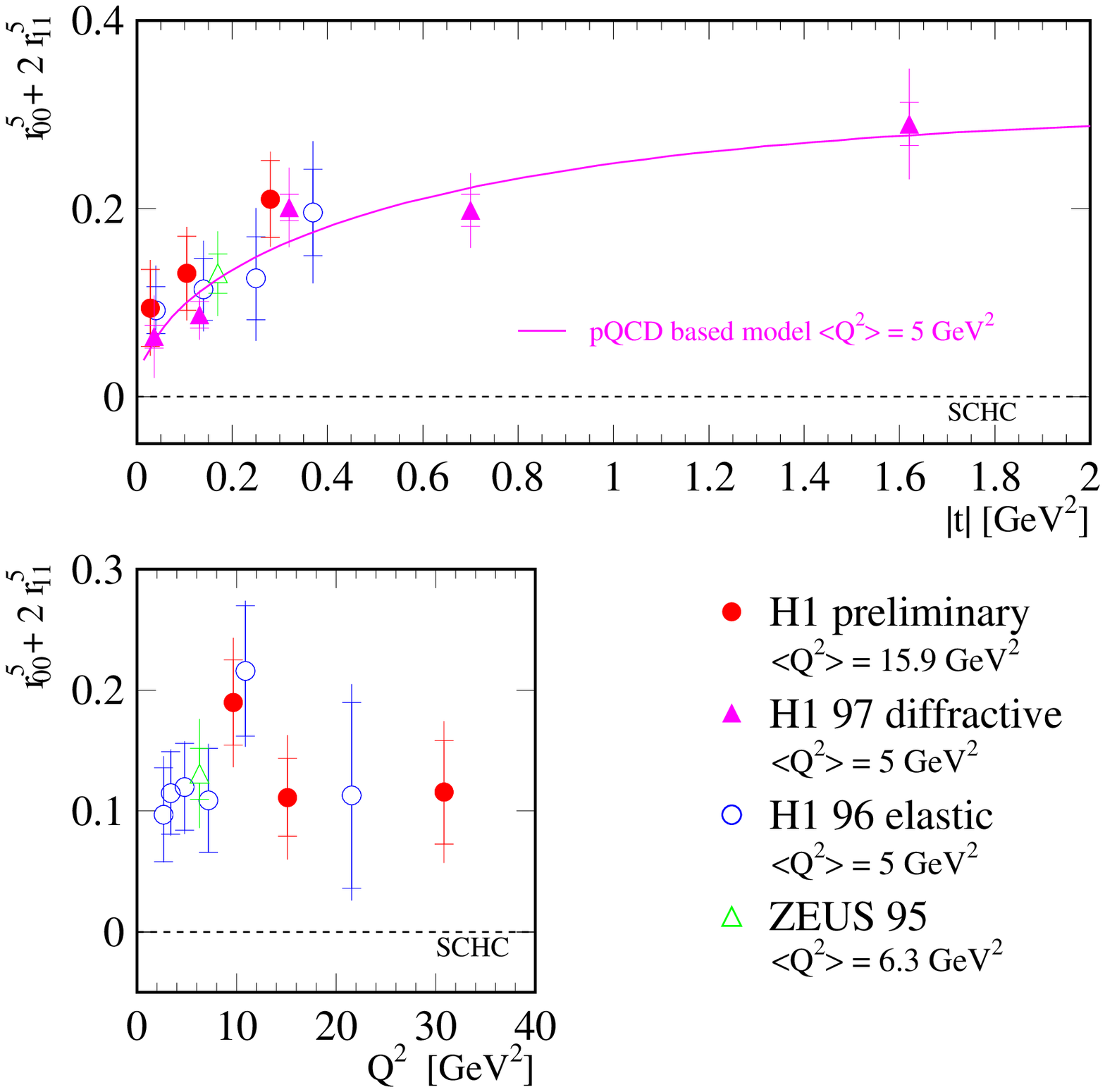,width=7cm,
      bbllx=10pt,bblly=10pt,bburx=520pt,bbury=520pt,clip= }
\end{minipage}
\par\noindent
\begin{minipage}[t]{7cm}
\hspace*{-0.2cm}
\epsfig{file=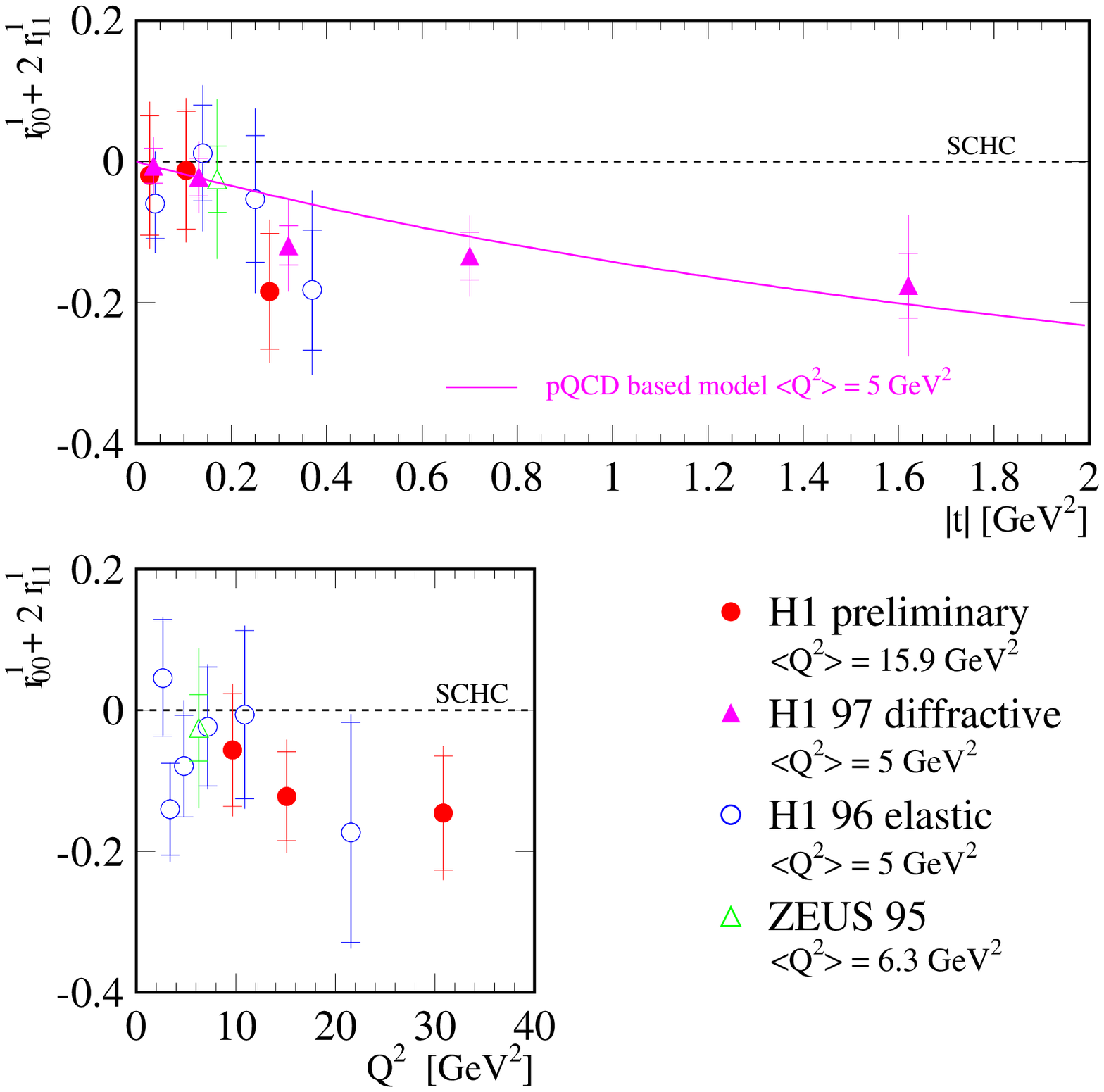,width=7cm,
      bbllx=10pt,bblly=10pt,bburx=520pt,bbury=520pt,clip=}
\par\noindent
Figure 8: Exclusive $\rho^{\circ}$ production. 
$|t|$ and $Q^2$-dependence of the SDME combination
$r^5_{00} + 2r^5_{11}$ (top) and 
$r^1_{00} + 2r^1_{11}$ (bottom). The curves show a pQCD  
model\cite{IvanovKirschner}, the dotted lines give the 
SCHC expectation.
\end{minipage} 
\end{minipage}
\vspace*{-16.5cm}
\par\noindent
\hspace*{7.3cm}
\begin{minipage}[t]{5.2cm}
Fig.~8 shows the determination of the SDME combinations $r^5_{00} + 2r^5_{11}$ 
and $r^1_{00} + 2r^1_{11}$, as functions of $|t|$ and $Q^2$. In the former
combination the element $r^5_{00}$ dominates. As seen, the previously 
observed significant SCHC
violation in this element
 persists at high values of $|t|$ and $Q^2$, and is increasing with 
$|t|$. The $\sim \sqrt{|t|}$ behaviour, which is predicted by pQCD for the 
combination $r^5_{00} + 2r^5_{11}$, describes the data well. \par\noindent
In the second
combination the element $r^1_{00}$ dominates. At large values of  
$|t|$ and $Q^2$ significant SCHC violation is observed. Also    
$r^1_{00}$ corresponds to the probability 
for a transverse photon to produce a longitudinal vector meson. Data are 
consistent with the predicted $\sim$linear $|t|$-dependence.
Note that a significant non-zero value for $r^1_{00}$ is not seen 
in the earlier studies
(Fig.~7) at low $|t|$ and  moderate $Q^2$ values.
\vspace*{0.1cm}
\par\noindent
The determination of the element $r^{04}_{00}$ from the angular distribution
$(4)$ is shown in Fig.~9, as function of $|t|$ and $Q^2$. The $|t|$-dependence
is constant, as predicted in pQCD. Note that data from two different $Q^2$
ranges are shown; $r^{04}_{00}$ is strongly increasing with $Q^2$.  
\end{minipage}
\end{minipage}
\par\noindent
The element $r^{04}_{00}$ represents the probability to produce a longitudinal
vector meson, from either a transverse or a longitudinal photon. 
It can be directly related to the ratio  
$R \ = \ \sigma_L/\sigma_T \ = \
1/\epsilon\,\cdot\,r^{04}_{00}/(1-r^{04}_{00})$. This relation is valid in
the SCHC approximation, i.e. if SCHC is assumed (in view of the previous 
findings, this is not quite correct and leads to a few~\%\ 
overestimate in $R$). The $R$ values determined from $r^{04}_{00}$  
are shown in Fig.~10 as a function of $Q^2$ and are well described  
by a pQCD calculation\cite{MRT}. For details of the deviation from the
naively expected linear $Q^2$ dependence, see \cite{MRT}.
\par\noindent
\begin{minipage}[t]{12.6cm}
\begin{minipage}[t]{7cm}
\hspace*{-0.2cm}
\epsfig{file=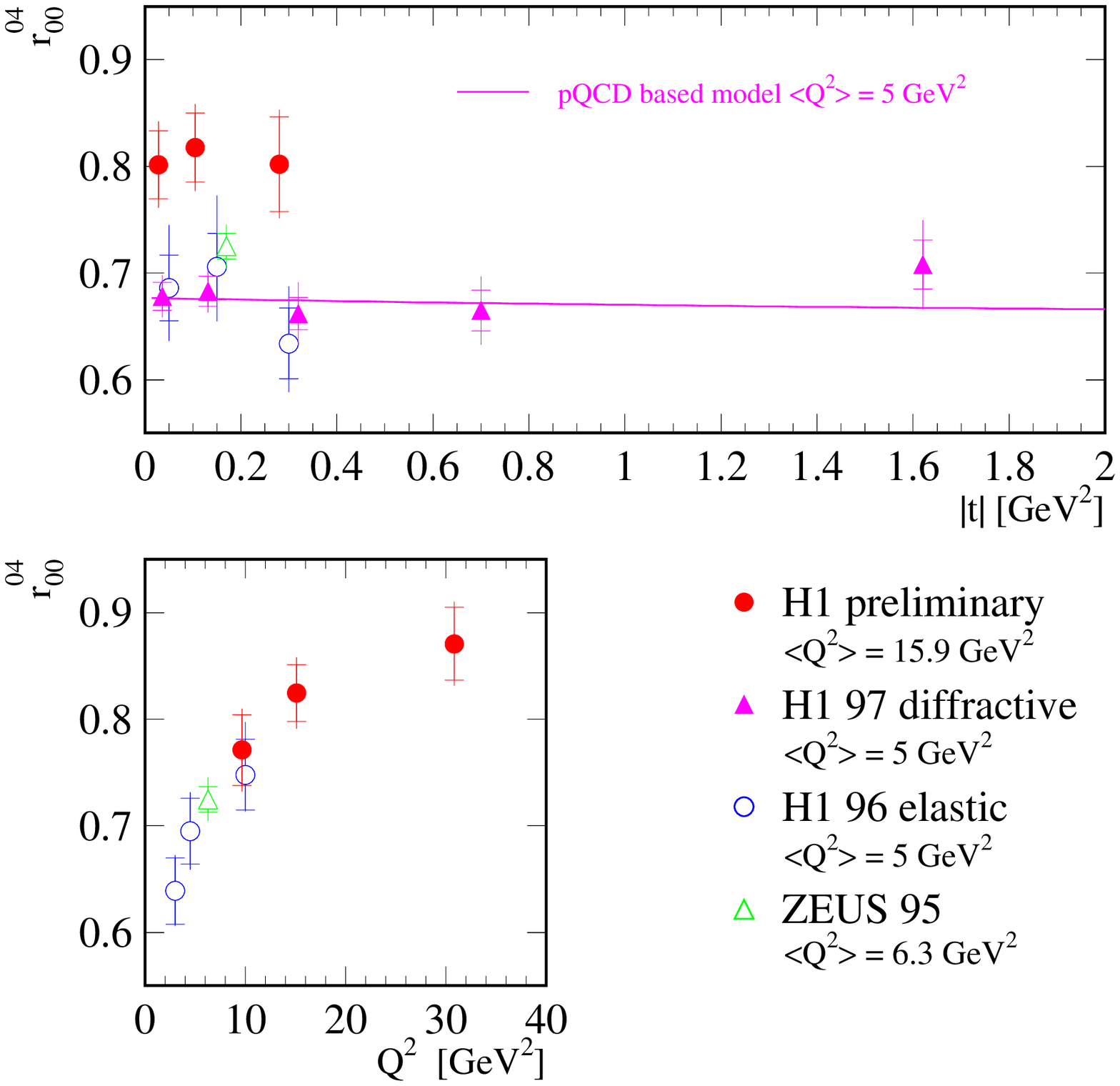,width=7cm,
      bbllx=10pt,bblly=10pt,bburx=520pt,bbury=520pt,clip=}
\vspace*{-0.3cm}
\par\noindent
Figure 9: $|t|$ and $Q^2$-dependence of the SDME 
$r^{04}_{00}$ in exclusive $\rho^{\circ}$ production. 
The curve shows the pQCD model \cite{IvanovKirschner}
prediction for the lower $Q^2$ range.
\end{minipage}
\vspace*{-9cm}
\par\noindent
\hspace*{7.5cm}
\begin{minipage}[t]{5cm}
\epsfig{file=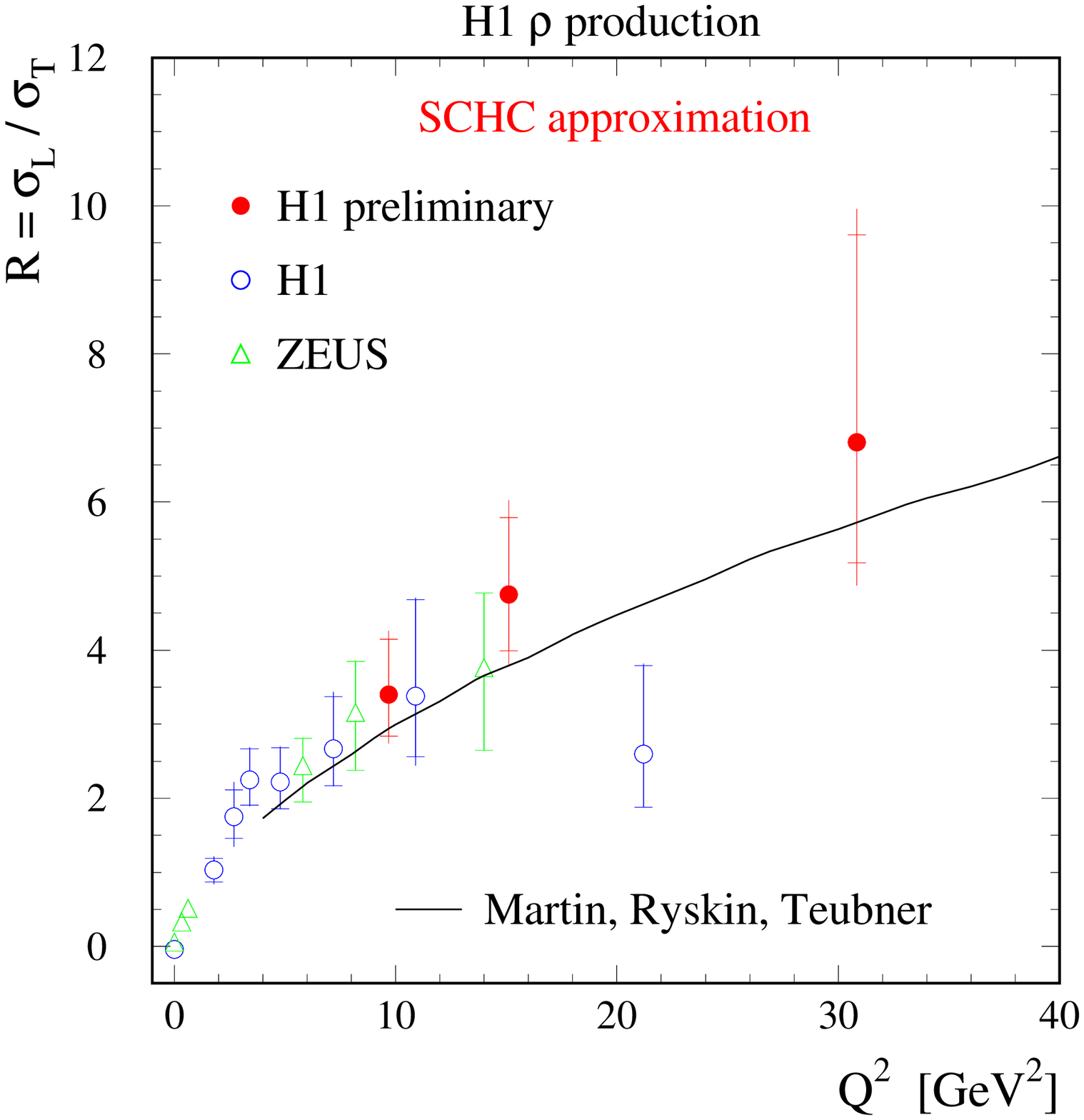,width=5.5cm} 
Figure 10:  Exclusive $\rho^{\circ}$ production. 
$Q^2$ dependence of the ratio $R = \sigma_L/\sigma_T$ in the 
SCHC approximation. The curve shows the pQCD model of \cite{MRT}. 
\end{minipage}
\end{minipage}
\vspace*{0.6cm}
\par\noindent
The examples given above have amply demonstrated that the quark mass,
 as well as $Q^2$ and $t$ all can serve as a hard scale
in pQCD calculations. The pQCD approach gives a successful description of 
exclusive vector meson production at large values of these scales, where the
traditional ``soft'' approach, using VDM and Regge theory, fails.
It should however be remembered that the HERA data not only confirm pQCD
calculations, but also inspire further development of the theory, in particular
where the data are NOT described; a recent example is given 
in \cite{newhightdata}, where the helicity structure of high $|t|$ 
photoproduction data is not reproduced by any of the available pQCD models.
\par\noindent
The coverage of the transition region in all involved scales makes HERA an
ideal place for these studies.
Much enlarged statistics for further advances in this field will soon be 
available, both in the remaining, still to be analyzed, HERA-I data, and from
the coming years of high luminosity running at the upgraded HERA-II. 
In particular it will then be interesting to see global analyses, 
using exclusive vector meson production data in combination with inclusive
measurement results, like
structure functions, jet analyses and inclusive charm production.
Such global analyses should yield the ultimate reachable 
precision and consistency in the 
parton densities and in the pQCD 
description of the various reactions, and eventually 
lead to a deeper understanding of the nature of diffraction.  

\section{Search for Odderon Exchange}

The developments in QCD in the last 20 years have led to the understanding
of Pomeron exchange as an exchange of two gluons, the simplest system for 
exchanging the quantum numbers of the vacuum. Similarly, the exchange of 
three gluons, also called Odderon exchange, is now recognized as an 
important and basic prediction of pQCD. After the early seminal papers
\cite{LUK73} established the Odderon 
as the $C=P=-1$ ``partner'' of the Pomeron,
several authors have suggested suitable reactions for its discovery,
based
upon asymmetries due to Pomeron-Odderon interference\cite{INTERF}, 
or exploiting the specific
quantum numbers of the Odderon exchange in exclusive production processes.
However, cross sections calculated in pQCD for such exclusive production are 
in general very small\cite{BAR2001}
and hardly accessible even with the HERA-II data.
\vspace*{0.2cm}
\par\noindent
\begin{minipage}[t]{12.6cm}
\begin{minipage}[t]{5cm}
\epsfig{file=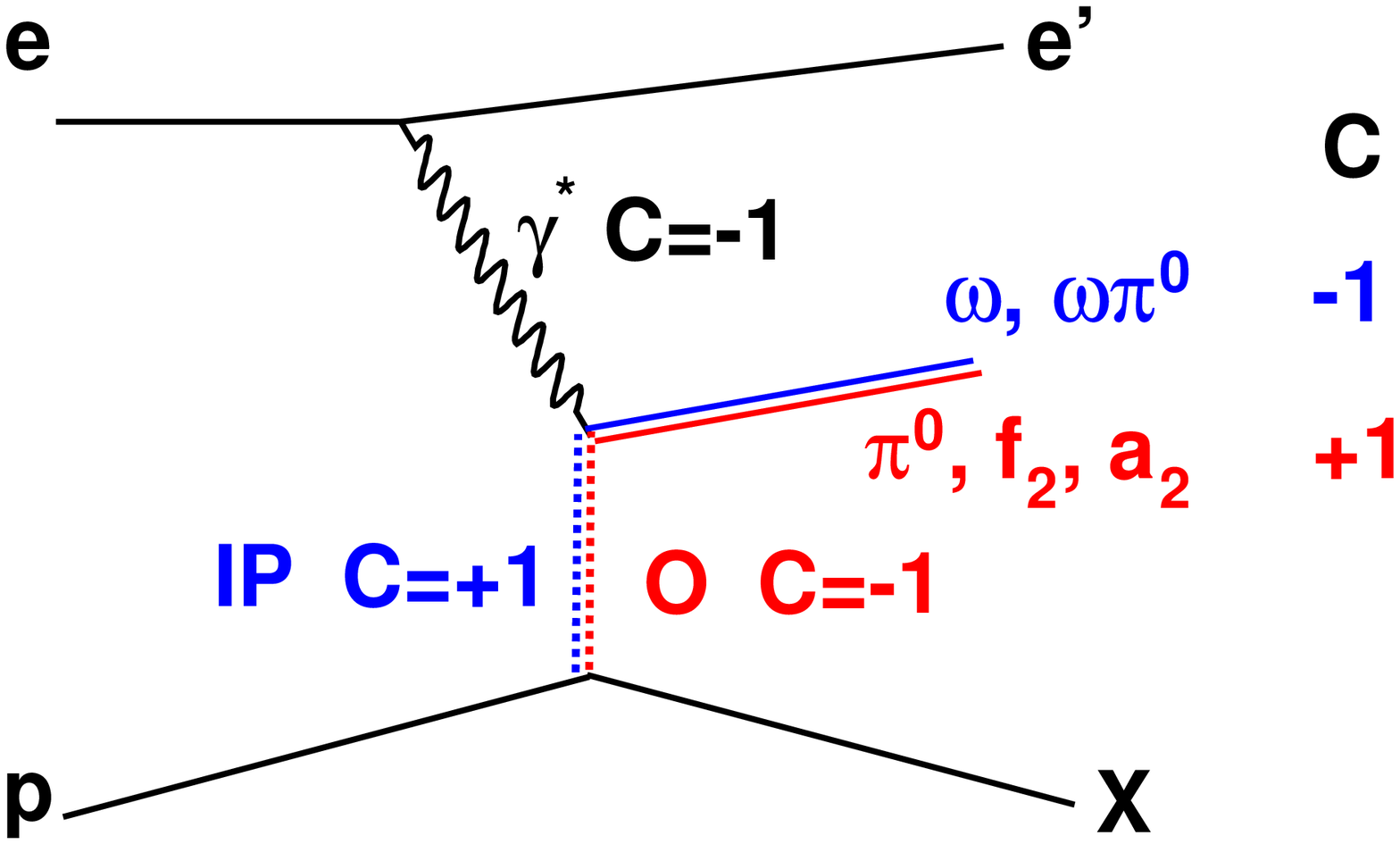,width=5cm}
Figure 11: Odderon-photon (Pomeron-photon) fusion, leading to 
C-parity $+1 \ \ (-1)$ exclusive final states. 
\end{minipage}
\vspace*{-5cm}
\par\noindent
\hspace*{5.5cm}
\begin{minipage}[t]{7cm}
\par\noindent
Recently, large cross sections for Odderon exchange in soft, exclusive
photoproduction processes have been predicted\cite{ERB99},
using a non-perturbative QCD approach based on the 
Stochastic Vacuum Model (SVM)\cite{DoschNachtmann}.
This model is very successful in describing a variety of 
data\cite{DoschDonnachie2002}, also data from
HERA. Its extension, including Odderon exchange in which the proton 
(a quark-diquark system in the model) is excited into a $P=-1\,\,N^*$
state, leads to predictions for the
exclusive photoproduc- 
\end{minipage}
\end{minipage}
\par\noindent
tion of pseudoscalars and tensor mesons,
reactions to which Pomeron exchange cannot contribute.
The predicted cross
sections are large enough to be seen at HERA, and H1 have made a search for
such reactions.
\par\noindent
The difference between Pomeron and Odderon exchange is obvious in the diagram
of Fig.~11. While Pomeron exchange results in $C=-1$ final states, Odderon
exchange leads to $C=+1$ final states. Multi-photon exclusive final 
states can then easily be separated into the two classes: 
even number of photons have  $C=+1$ (Odderon exchange), 
odd number of photons $C=-1$ (Pomeron exchange). 
Candidate multi-photon final states
are e.g. $\pi^{\circ} \rightarrow  2 \gamma, \ 
f_2(1270)  \rightarrow \pi^{\circ}\pi^{\circ} 
           \rightarrow  4 \gamma$ and
$a_2(1320) \rightarrow  \pi^{\circ}\eta 
             \rightarrow  4 \gamma$,
all with $C=+1$, and
$\omega \rightarrow \gamma \pi^{\circ}  
             \rightarrow 3 \gamma$ 
and
$b_1(1235) \rightarrow  \omega \pi^{\circ} 
             \rightarrow   5 \gamma$, both with $C=-1$.
\vspace*{0.1cm}
\par\noindent
The photons were detected in the backward calorimeters of H1 and the scattered
electron in the small angle electron tagger (tagged photoproduction). The 
$3\gamma$ mass spectrum\cite{BUD01} 
for the exclusive three-photon sample is shown in 
Fig.~12. Only events which are candidates for the final state 
$\gamma\pi^{\circ}$ are included. A clear $\omega$ peak is seen above the 
background. The preliminary cross section (at $W_{\gamma p}\approx 200$ Gev),
$ \sigma(\gamma p \rightarrow \omega p) =
                   (1.25 \pm 0.17 \pm 0.22) \ \mu{\rm b}$     
agrees very well 
with the expectation from the $W^{\delta}$ dependence for $\omega$ 
photoproduction (see Fig.2).
\par\noindent
\begin{minipage}[t]{12.6cm}
\begin{minipage}[t]{6cm}
\epsfig{file=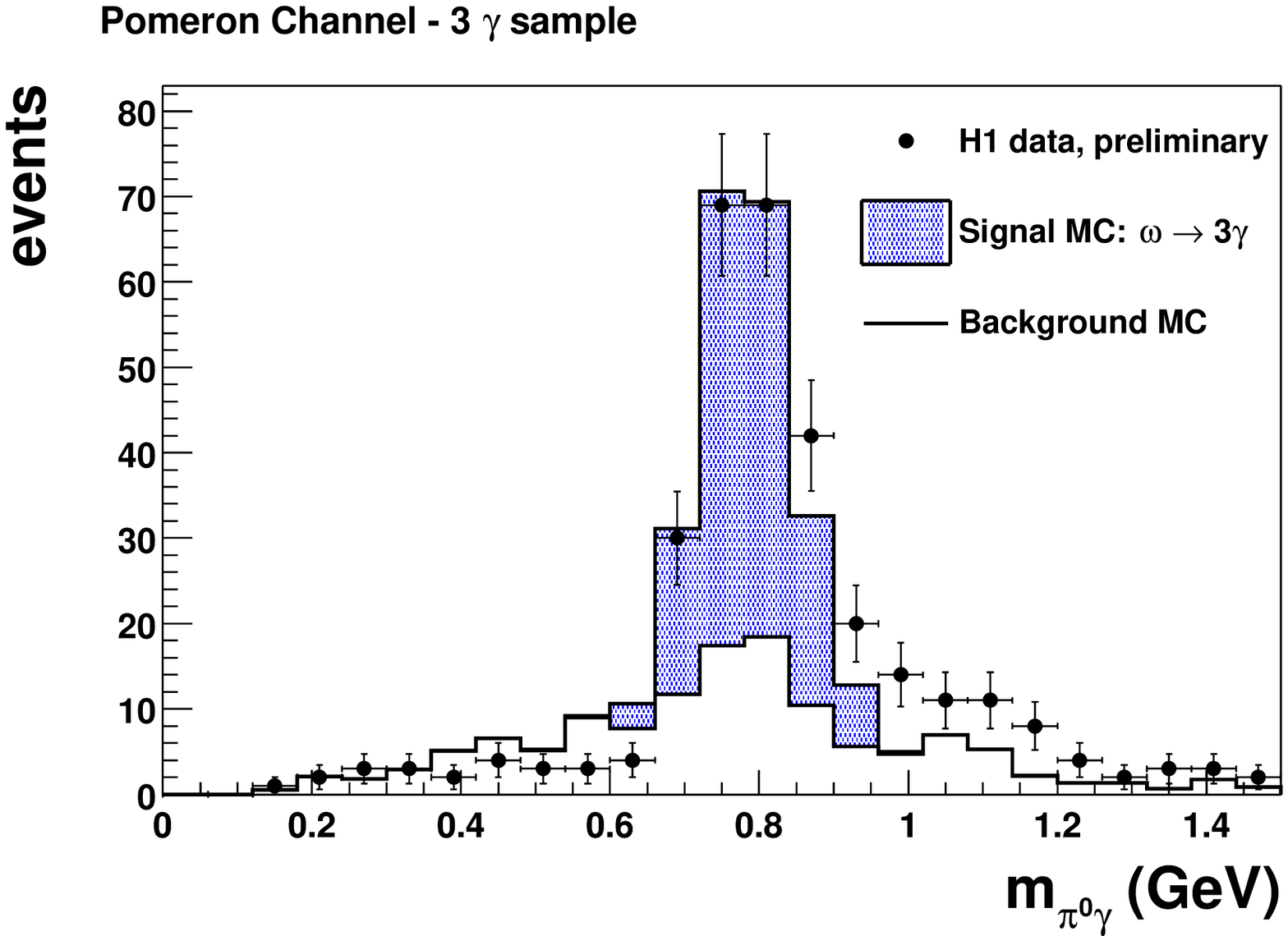,width=6cm}
Figure 12: Three photon invariant mass, for $\gamma\pi^{\circ}$ event
candidates. Model prediction and expected background (hatched and white
histograms) are also shown.
\end{minipage}
\vspace*{-6.3cm}
\par\noindent
\hspace*{6.5cm}
\begin{minipage}[t]{6cm}
\epsfig{file=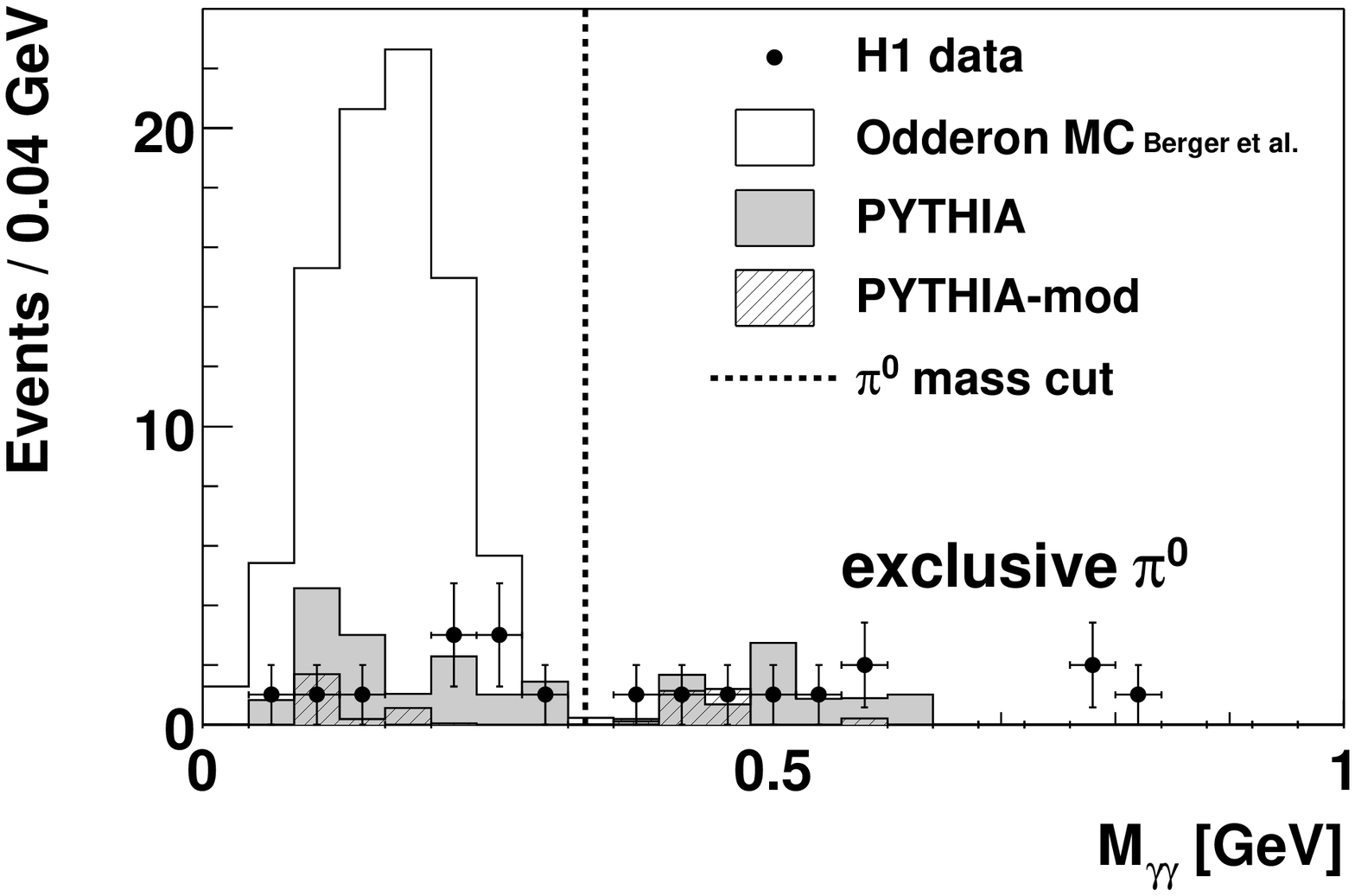,width=6cm,
      bbllx=0pt,bblly=0pt,bburx=560pt,bbury=375pt,clip= }
Figure 13: Two photon invariant mass, for exclusive two photon event
candidates. The expectations from background and model are shown as
hatched and white histograms. 
\end{minipage}
\end{minipage}
\vspace*{0.2cm} 
\par\noindent
The measurements in the 3-photon and and also 5-photon final states 
($\gamma p \rightarrow \omega\pi^{\circ} X$, not discussed here) are 
interesting in themselves, but for the present purpose can be taken as
proof that the photon and $\pi^{\circ}$ detection with the H1 detector 
is well understood. Turning now to the $C=+1$ multi-photon final states,
the $2\gamma$ mass spectrum of exclusive two-photon 
events\cite{H1Odderonpi0} is shown in 
Fig.~13. Only very few events are seen, compatible with the expected 
background, and there is no indication of 
a $\pi^{\circ}$ peak. This is in stark contrast to the expectation from 
the model prediction, $>100$ events. Taking all events below the generous
$\pi^{\circ}$ mass cut as signal events, the upper limit
\hspace*{1cm} 
$
\sigma (\gamma p
       \rightarrow^{\hspace*{-2ex}\mathbb{O}} \ \pi^{\circ}N^{\ast}) < 49 \ 
             {\rm nb}   \hspace*{0.5cm} (95 \%\ {\rm CL})
$ \newline
is derived, for $<W>=215$ GeV and the range $0.02<|t|<0.3$ GeV$^2$ covered by
the experiment. The limit is clearly below the predicted cross section 
of $>200$ nb. Note that the limit is given for the $N^*$ final state, since
in this case a neutron from the $N^*$ decay was detected and used
in the trigger.
\par\noindent
$4\gamma$ mass distributions\cite{BUD01} 
are shown for $\pi^{\circ}\pi^{\circ}$ and 
 $\pi^{\circ}\eta$ final state candidate events in Figs.14a and 14b, 
respectively\cite{BUD01}. 
Again, the data are consistent with the expected background,
and lie below the model prediction for exclusive 
$f_2(1270)$ and $a_2(1320)$ resonance production via Odderon exchange. The 
(still preliminary) upper limits are in these cases (95 \%\ CL) 
\newline \hspace*{1cm}
$
\sigma(\gamma p\to^{\hspace*{-2ex}\mathbb{O}}\
       f_2(1270)X) < 16~{\rm nb} \ \ {\rm and} \ 
\sigma(\gamma p\to^{\hspace*{-2ex}\mathbb{O}}\
       a_2(1320)X) < 96~{\rm nb}, 
$
\newline
to be compared to the model predictions of 21 and 190 nb, respectively. 
\vspace*{0.3cm}    
\par\noindent
\begin{minipage}[t]{12.6cm}
\begin{minipage}[t]{6cm}
\epsfig{file=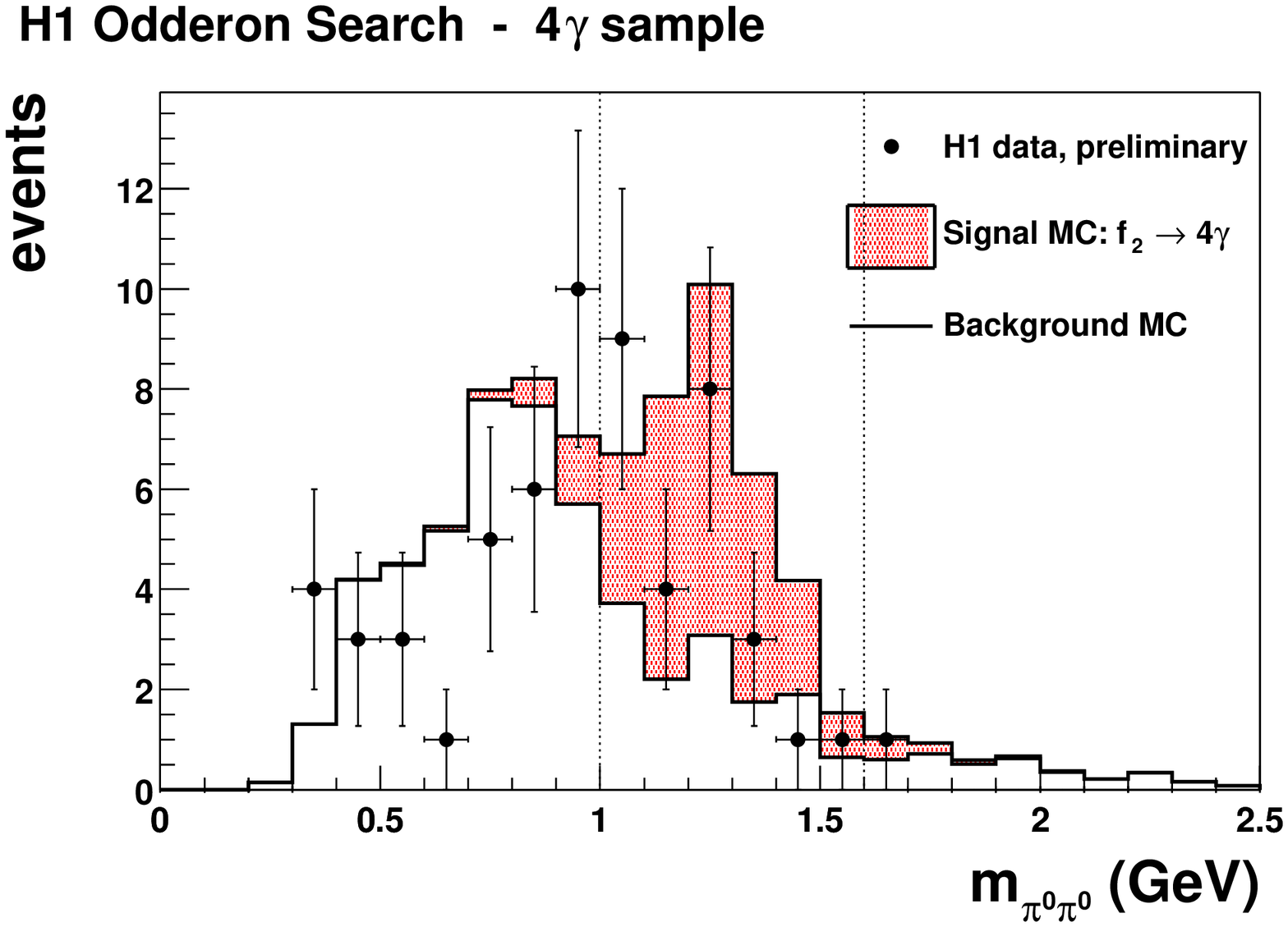,width=6cm}
\end{minipage}
\vspace*{-4.3cm}
\par\noindent
\hspace*{6.5cm}
\begin{minipage}[t]{6cm}
\epsfig{file=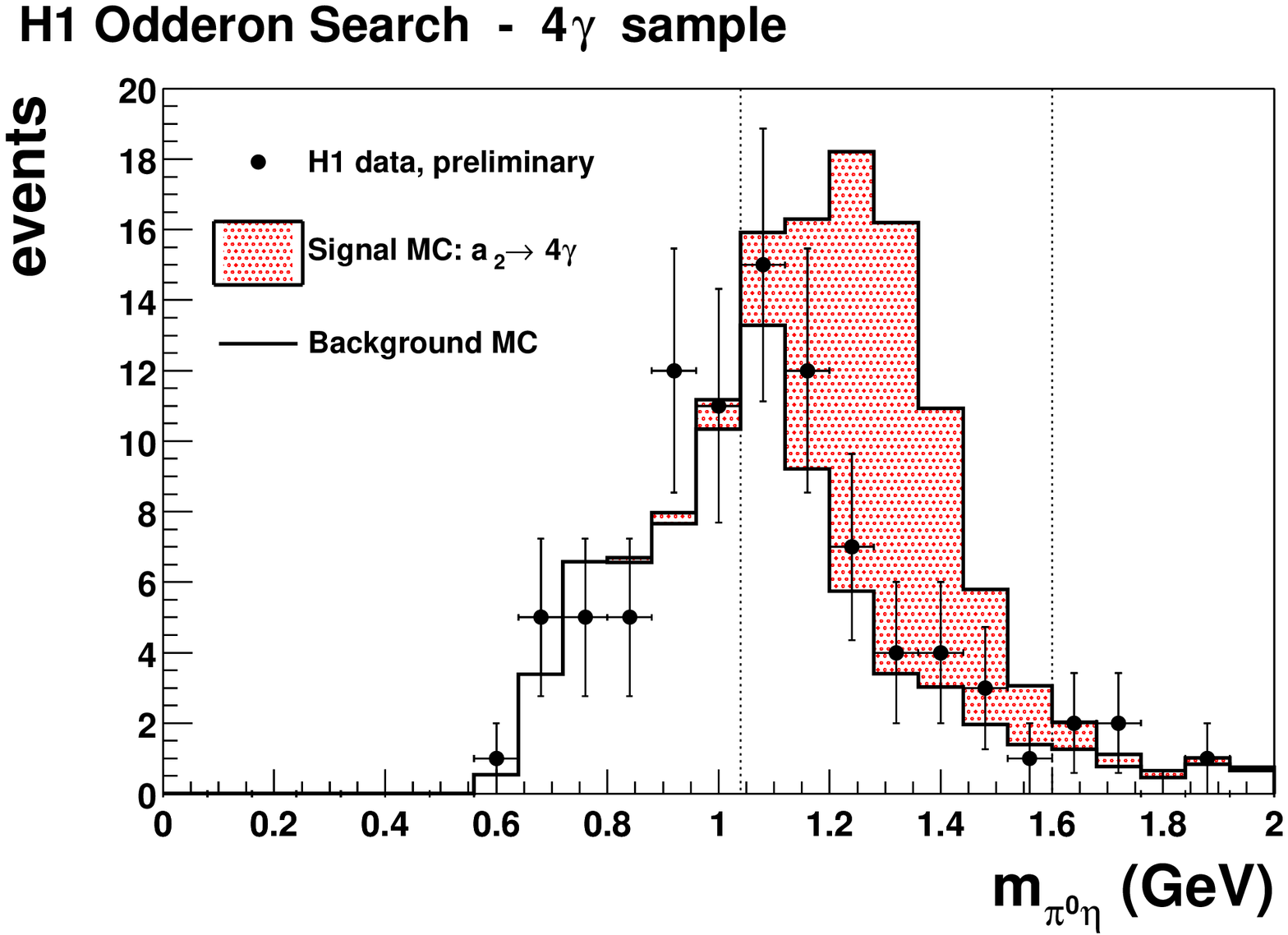,width=6cm}
\end{minipage}
\par\noindent
Figure 14: Four photon invariant mass, for a) $\pi^{\circ}\pi^{\circ}$ and
b) $\pi^{\circ}\eta$ event candidates. Also shown are the expectations from
model and background (hatched and white histograms). Vertical lines indicate
the $f_2$ and $a_2$ mass regions for the upper limits derivation.
\end{minipage}
\vspace*{-4cm}
\par\noindent
\hspace*{4.9cm} a) \hspace*{2.2cm} b)
\vspace*{3.8cm}
\par\noindent
Thus, no evidence of Odderon exchange was found at the cross section
levels predicted by the Stochastic Vacuum Model. This result is currently
not understood. A possible explanation could be that the Odderon intercept
$\alpha_{\mathbb{O}}$ is much smaller than unity;
the SVM predictions are made for $W=20$~GeV and $\alpha_{\mathbb{O}}\gsim~1$.
Indeed,
$\alpha_{\mathbb{O}}<0.7$ is compatible with the upper limit for exclusive
$\pi^{\circ}$ photoproduction and also with alternative 
predictions\cite{Kaidalov99}.
\par\noindent
Nevertheless, the search at HERA for the Odderon will be continued, 
both with higher
statistics for the above reactions (noting that the upper limit for
$f_2(1270)$ production is just
below the model prediction), and in other final states where \eg predictions
of asymmetry due to Pomeron-Odderon interference\cite{INTERF} can be tested. 

\section{Open Charm Spectroscopy}

The study of inclusive heavy quark production at HERA has many facets, see
\eg the extensive review in \cite{Busseyhq}.
In LO QCD, charm quarks are produced via the Boson-Gluon-Fusion
(BGF) diagram in Fig.~15a. The photon interacts pointlike. Taking photon 
structure into account (``resolved'' photon), the diagrams in Figs.~15b and 
15c contribute, with either a gluon or a $c$ quark interacting with the 
gluon from the proton structure. In the latter case (``charm excitation'')
the photon remnant will also contain a charm quark. Thus, charm production
gives access both to the gluon density of the proton (direct sensitivity 
via BGF) and to the charm content (charm structure function) of the proton 
and photon. The $c$ quark mass together with $Q^2$ provide hard scales for 
pQCD calculations of charm production, and the HERA data are used to test 
these calculations.
\vspace*{0.3cm}
\par\noindent
\begin{minipage}[t]{12.6cm}
\begin{minipage}[t]{3cm}
\epsfig{file=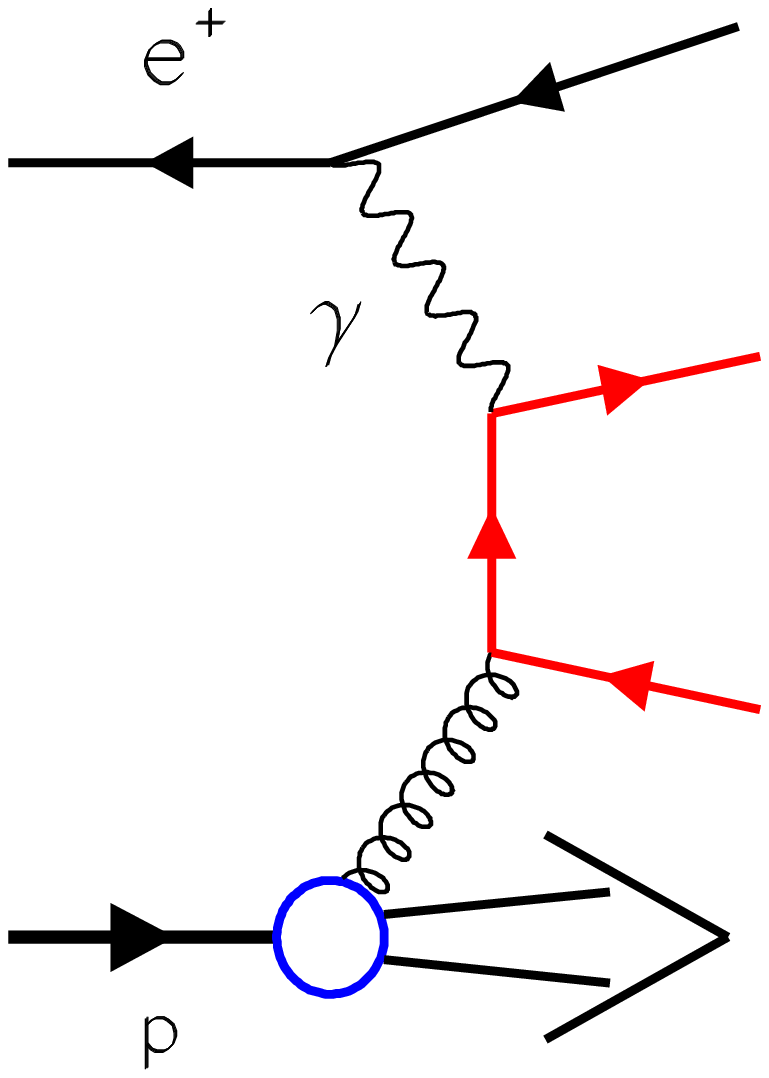,width=3cm,
      bbllx=125pt,bblly=130pt,bburx=360pt,bbury=445pt,clip= }
\end{minipage}
\vspace*{-4cm}
\par\noindent
\hspace*{4.4cm}
\begin{minipage}[t]{3.3cm}
\epsfig{file=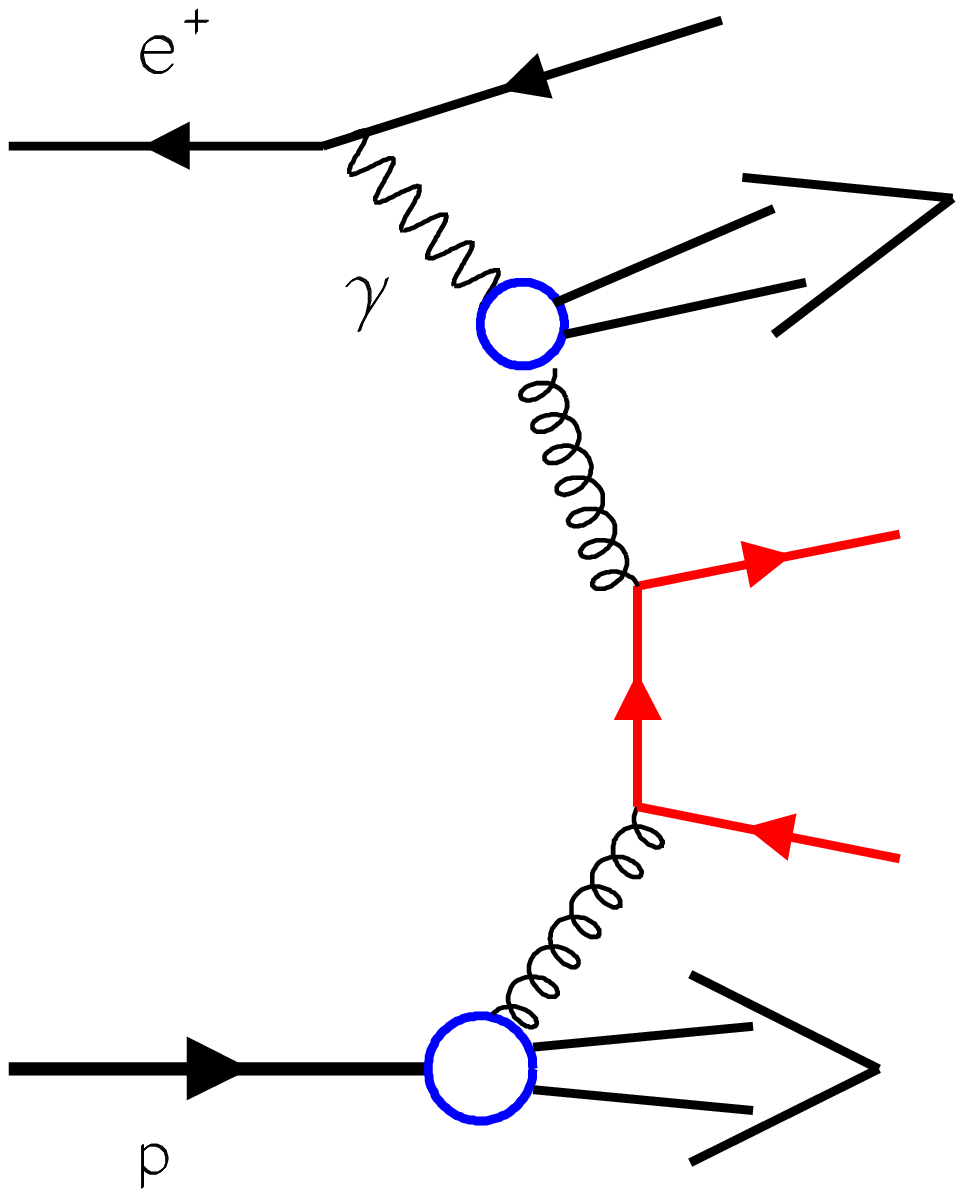,width=3.3cm,
      bbllx=110pt,bblly=110pt,bburx=400pt,bbury=460pt,clip= }
\end{minipage}
\vspace*{-4.2cm}
\par\noindent
\hspace*{8.5cm}
\begin{minipage}[t]{3.8cm}
\epsfig{file=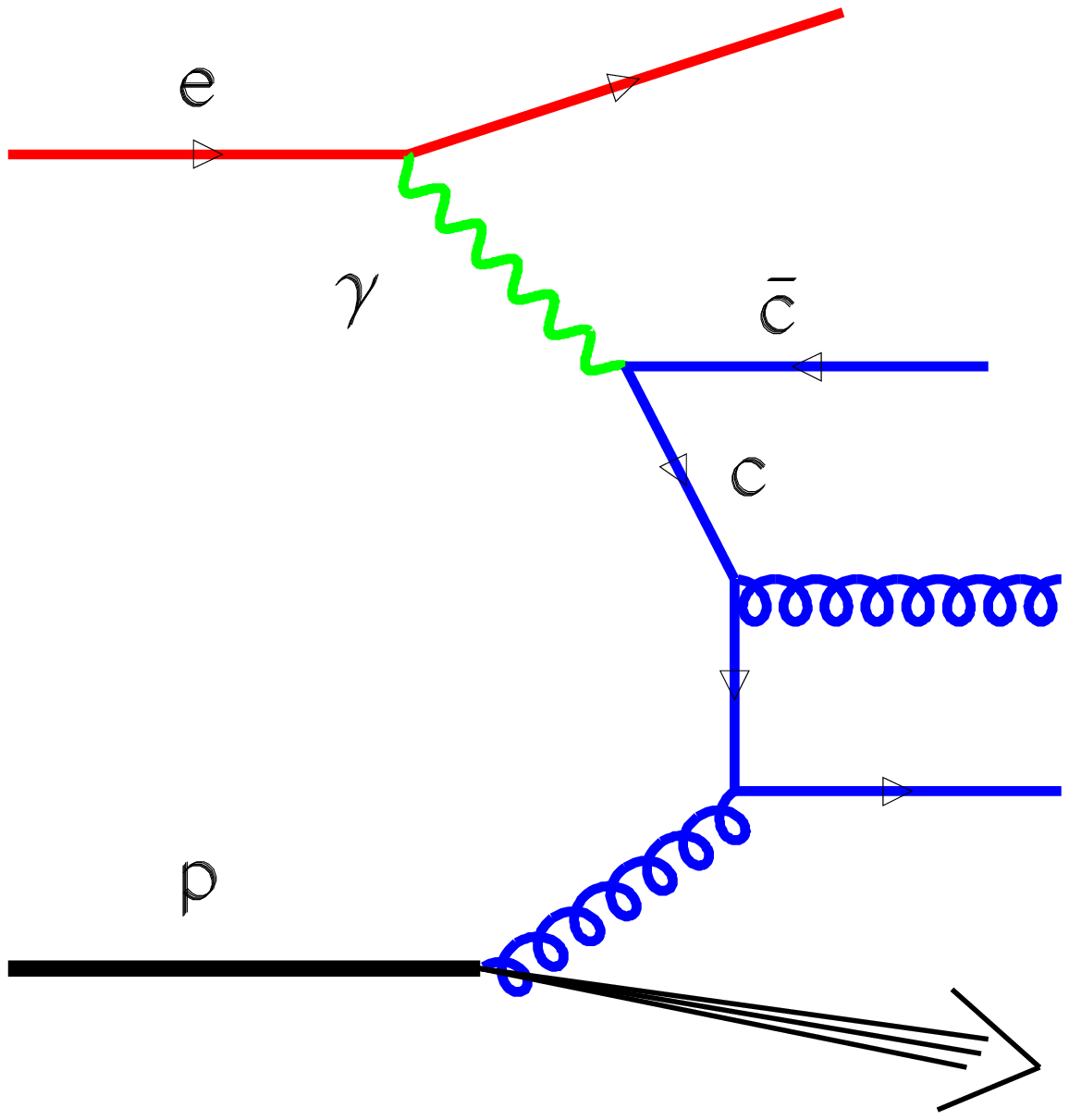,width=3.8cm,
      bbllx=50pt,bblly=135pt,bburx=400pt,bbury=490pt,clip= }
\end{minipage}
\vspace*{-2cm}
\par\noindent
a) \hspace*{4cm} b) \hspace*{3.8cm} c)
\vspace*{1.8cm}
\par\noindent
Figure 15: Diagrams contributing to charm production in LO QCD. \newline 
a) boson-gluon fusion, b) and c) resolved photon interactions.
\end{minipage}
\vspace*{0.3cm}
\par\noindent
However, the fragmentation of the charm quarks, once produced, cannot be
predicted in pQCD. Non-perturbative, empirical models of the 
charm fragmentation can be tuned with the measurements at HERA and elsewhere, 
and by comparison of HERA $ep$ data with data from $e^+e^-$, $\gamma\gamma$
and $p\bar{p}$ collisions the principle of universality of 
charm fragmentation can be tested.
Thus, charm spectroscopy is an important aspect of heavy quark physics at 
HERA; here, some recent results from the H1 and ZEUS
measurements of D-meson production
are presented.
\vspace*{0.3cm}
\par\noindent
\begin{minipage}[t]{12.6cm}
\begin{minipage}[t]{6cm}
\epsfig{file=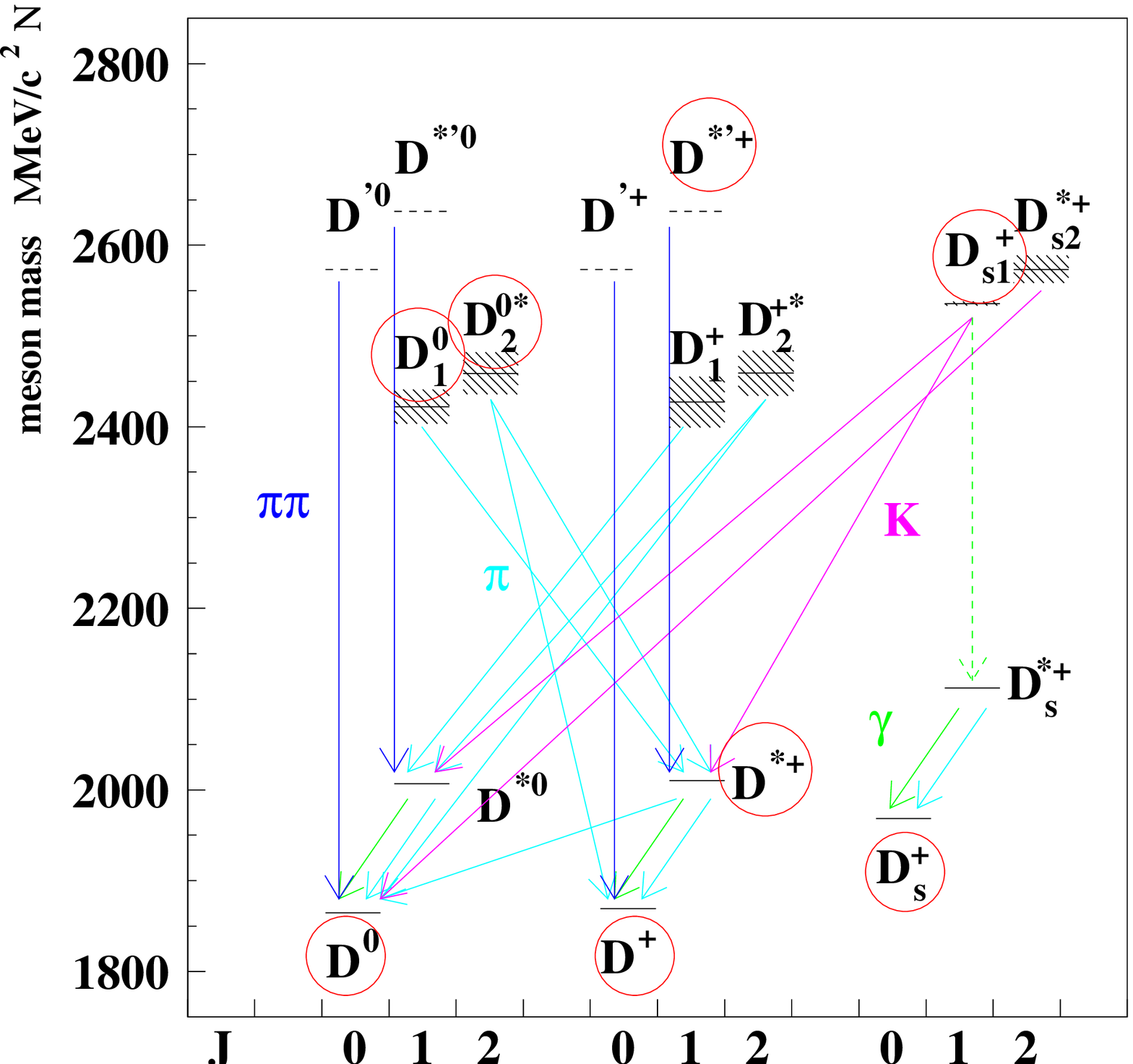,width=6cm}
Figure 16: 
D-meson spectroscopy. The rings indicate the states studied so far by H1 and
ZEUS.
\end{minipage}
\vspace*{-7.1cm}
\par\noindent
\hspace*{6.4cm}
\begin{minipage}[t]{6cm}
The spectroscopy of $D$-mesons is shown in Fig.16. The states which are ringed
have so far been addressed in the HERA studies. 
The access to these states is so far limited to decay modes into final states
consisting of charged particles only.
They include the 
$L=0$ states, namely the pseudoscalar $D$-mesons and the vector $D^*$-mesons,
as well as the heavier $P$-wave states. A search has also 
been made for radially excited $D^*$-states.
\par\noindent
The experimental detection is based on two different techniques, either a
lifetime measurement, using a high 
\end{minipage}
\end{minipage}
\par\noindent
precision vertex detector to locate
the charmed meson decay vertex, or
the well-known $\Delta m$ technique, utilizing the limited phasespace 
(mass difference $\sim~10$ MeV) for 
the decay $D^*(2010) \rightarrow D\pi_s$, where $\pi_s$ is a ``slow'' pion.
\par\noindent
{\bf \boldmath$D$-meson production cross sections: \ } New measurements of
the production cross sections for all pseudoscalar D-mesons (the data are 
shown in Fig.~17), as well as for
the $D^*(2010)$, are available from H1\cite{NewH1charmdata}. 
They are based on the H1 silicon vertex detector (CST) data and the 
$D^+$ measurement\footnote{The charge conjugate states are always implicitly
included.} is the first at HERA.
\vspace*{0.2cm}
\par\noindent
\begin{minipage}[t]{12.6cm}
\begin{minipage}[t]{4cm}
\epsfig{file=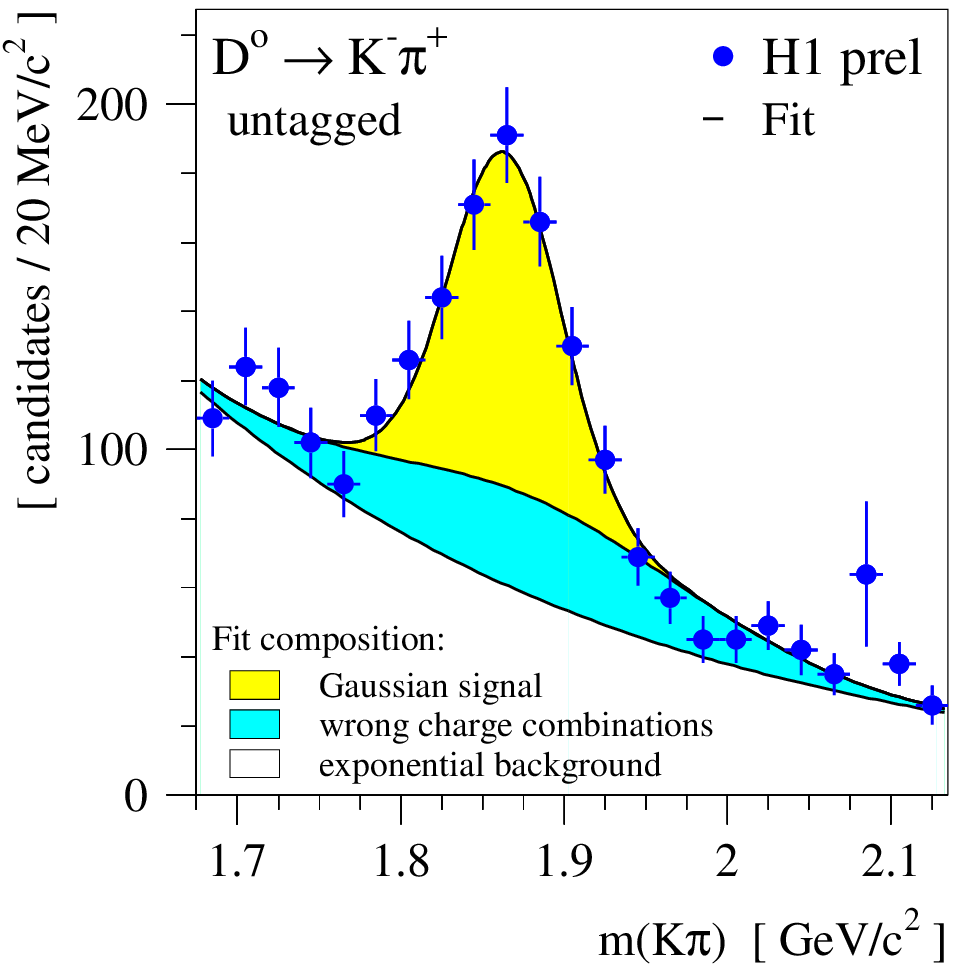,width=4cm,
        bbllx=0pt,bblly=0pt,bburx=285pt,bbury=285pt,clip= }
\par\noindent
\epsfig{file=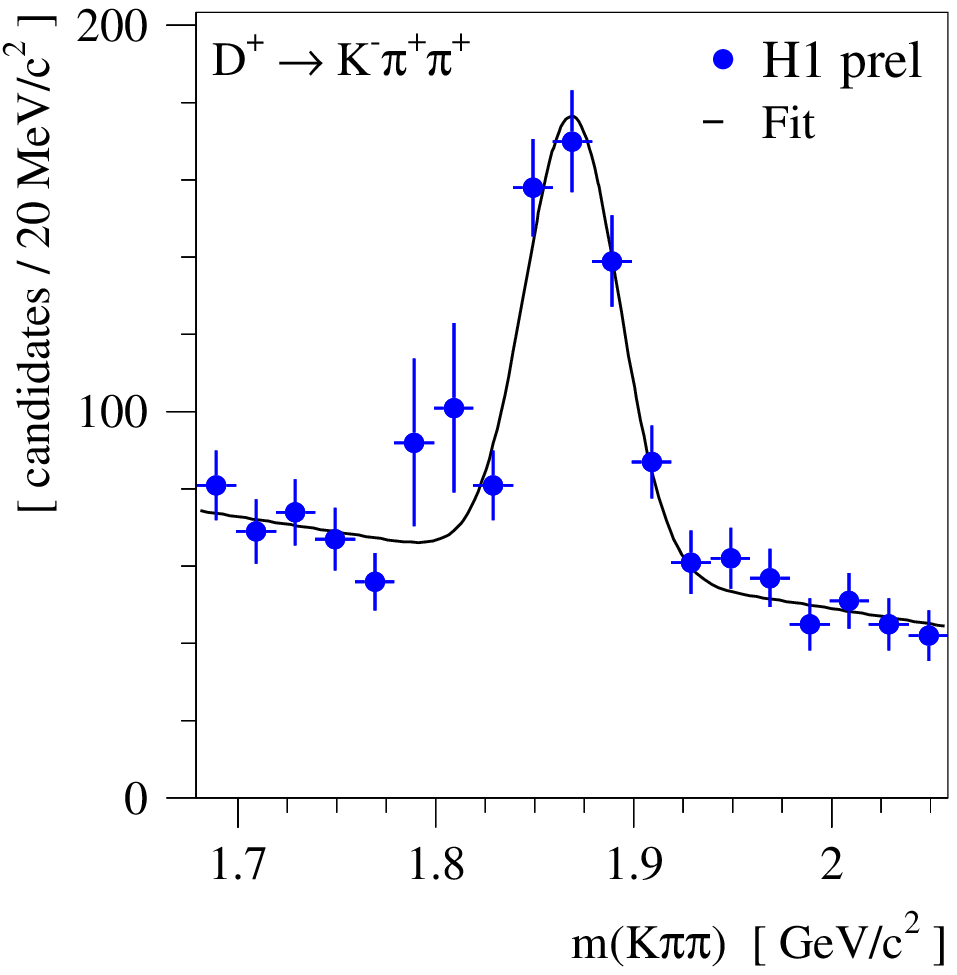,width=4cm,
        bbllx=0pt,bblly=0pt,bburx=285pt,bbury=285pt,clip= }
\end{minipage}
\vspace*{-8cm}
\par\noindent
\hspace*{4.5cm}
\begin{minipage}[t]{8cm}
\epsfig{file=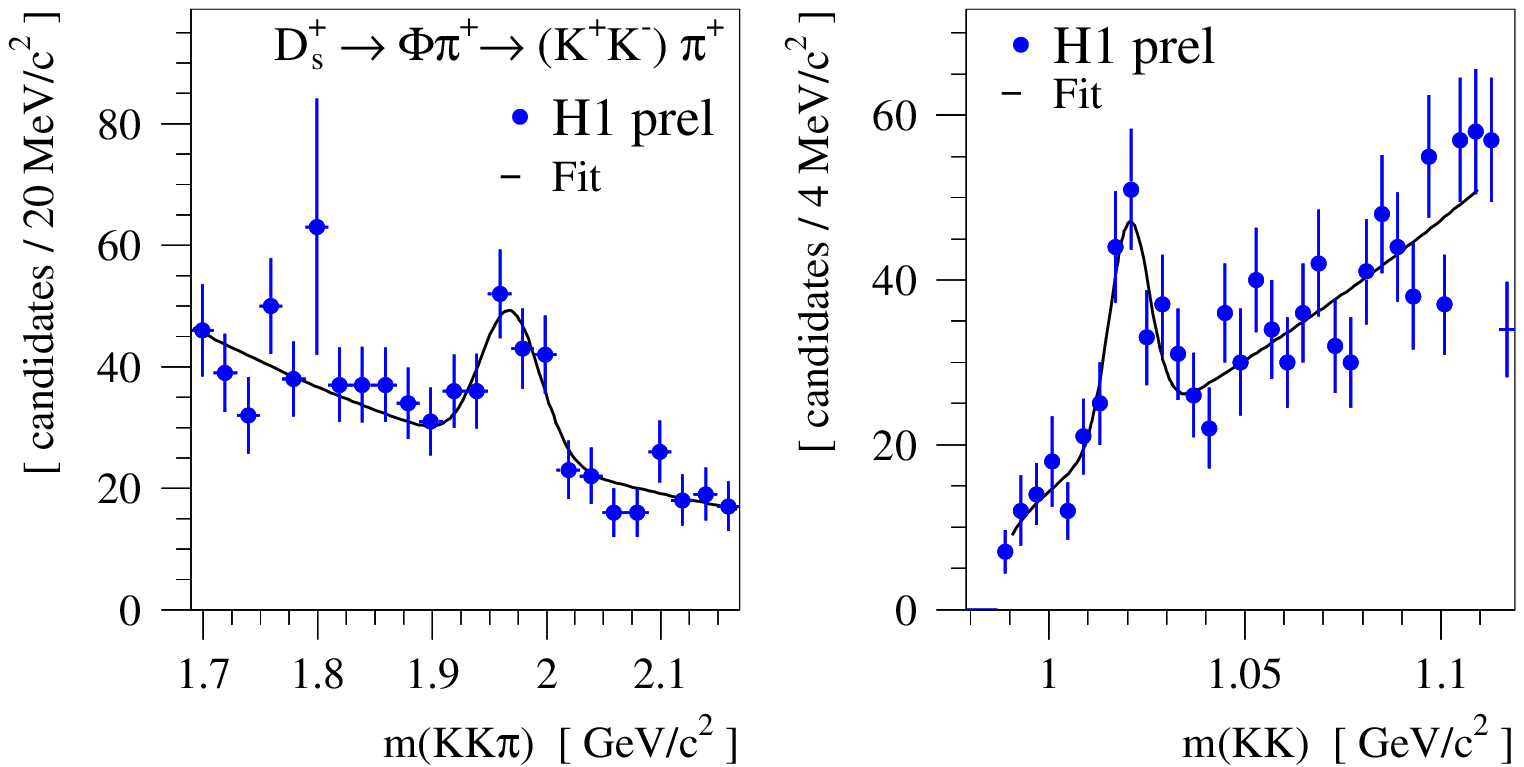,width=8cm,
        bbllx=0pt,bblly=0pt,bburx=445pt,bbury=225pt,clip= }
\end{minipage}
\vspace*{0.3cm}
\par\noindent
\hspace*{5cm}
\begin{minipage}[t]{7.5cm}
Figure 17: Inv.mass distributions of a) $K^{\mp}\pi^{\pm}$, 
b) $K^{\mp}\pi^{\pm}\pi^{\pm}$, c) $\phi\pi^{\pm}\rightarrow K^+K^-\pi^{\pm}$ 
and d) $K^+K^-$. 
Curves show fits of Gaussian signals $+$ background. H1 preliminary data
using the CST vertex detector.
\end{minipage}
\end{minipage}
\vspace*{-6.5cm}
\par\noindent
\hspace*{3cm} a)
\vspace*{1.3cm}
\par\noindent
\hspace*{6cm} c) \hspace*{5cm} d)
\vspace*{2cm}
\par\noindent
\hspace*{3cm} b)
\vspace*{2.6cm}
\par\noindent
\begin{minipage}[t]{12.6cm}
\begin{minipage}[t]{6.6cm}
 \begin{tabular}{|c|c|c|}
  \hline
   H1 & \multicolumn{2}{c|}{Visible Cross Section (nb)} \\ 
   \cline{2-3}
   Prel. & Measurement  &  LO QCD \\
\hline
   \dc  & 2.16 $\pm 0.19$ $^{+0.46}_{-0.35}$ &  2.45 $\pm 0.30$ \\
   \dn  & 6.53 $\pm 0.49$ $^{+1.06}_{-1.30}$ &  5.54 $\pm 0.69$ \\ 
   \ds  & 1.67 $\pm 0.41$ $^{+0.54}_{-0.54}$ &  1.15 $\pm 0.30$ \\ 
   \dst & 2.90 $\pm 0.20$ $^{+0.58}_{-0.44}$ &  2.61 $\pm 0.31$ \\
  \hline 
 \end{tabular}
\vspace*{0.2cm}
\par\noindent
Table 1: Visible cross sections for $D$-meson production, compared
to the LO QCD (AROMA 2.2) prediction.
\end{minipage}
\vspace*{-4.6cm}
\par\noindent
\hspace*{7cm}
\begin{minipage}[t]{5.5cm}
The measured visible cross sections are given in Table 1.
The kinematic range of the measurements is \newline
\hspace*{0.3cm} $2<Q^2<100$ GeV$^2$, \newline
\hspace*{0.3cm} $0.05<y<0.7$, \newline
\hspace*{0.3cm} $p_{\perp}(D)>2.5$ GeV \ \ \ and \newline
\hspace*{0.3cm} $|\eta(D)|<1.5$, \newline
where $y$ is the inelasticity and $\eta$ the pseudo-rapidity.
\end{minipage}
\end{minipage}
\vspace*{0.3cm}
\par\noindent
The cross sections in Table 1 are compared to the Leading Order (LO)
QCD prediction, obtained with the Monte Carlo (MC) generator 
program AROMA~2.2\cite{AROMA}. Good agreement with the data
is observed. Note that both the measured and the predicted cross sections
include contributions from decays of $b$ flavoured hadrons into $D$-mesons.
These contributions are taken from the AROMA simulation and the cross 
section measurements are converted into fragmentation factors 
$f(c\rightarrow D)$, using the expression
$$
f^{}(c\rightarrow D) = \frac{\sigma^{meas}_{vis}(ep \rightarrow eDX) - \sigma^{MC}_{vis} (ep \rightarrow b\bar{b} \rightarrow eDX)} {\sigma^{MC}_{vis} (ep \rightarrow c\bar{c} \rightarrow eDX)} \cdot { f_{w.a.}(c \rightarrow D)},
$$
where the fragmentation dependence $f_{w.a.}$ (the world averages of the 
charm fragmentation factors are used in the hadronisation part of the
MC simulation) has been removed. The resulting measured fragmentation 
factors are given in Table 2 and compare well with the world average 
values\cite{Gladilin99},
which are dominated by the LEP $e^+e^-$ results.
\vspace*{0.3cm}
\par\noindent
\begin{minipage}[t]{12.6cm}
\begin{minipage}[t]{9cm}
 \begin{tabular}{|c|c|c|}
  \hline
   H1 & \multicolumn{2}{c|}{Fragmentation Factor}    \\
    \cline{2-3}
   Prel.  & Measurement   &  World Average  \\ 
  \hline
   \dc  & 0.202 $\pm 0.020$ $^{+0.045}_{-0.033}$ $^{+0.029}_{-0.021}$ 
        & 0.232 $\pm 0.018$        \\
   \dn  & 0.658 $\pm 0.054$ $^{+0.115}_{-0.148}$ $^{+0.086}_{-0.048}$   
        & 0.549 $\pm 0.026$          \\
   \ds  & 0.156 $\pm 0.043$ $^{+0.036}_{-0.035}$ $^{+0.050}_{-0.046}$  
        &  0.101 $\pm 0.027$         \\
   \dst & 0.263 $\pm 0.019 $ $^{+0.056}_{-0.042}$ $^{+0.031}_{-0.022}$    
        & 0.235 $\pm 0.010$          \\   
  \hline 
 \end{tabular}
\vspace*{0.2cm}
\par\noindent
Table 2: Measured charm fragmentation factors compared with the world 
average values\cite{Gladilin99}. 
\end{minipage}
\vspace*{-4.2cm}
\par\noindent
\hspace*{9.3cm}
\begin{minipage}[t]{3.3cm}
Since the cross section measurements are all made in the same
kinematic range, ratios of the fragmentation factors can be used to 
calculate several interesting quantities, characterizing the 
\end{minipage}
\end{minipage}
\par\noindent
fragmentation. Thus the fraction of vector mesons $P_V$ produced in the
fragmentation, the ratio $R_{u/d}$ of $u$ and $d$ quarks participating in 
the charm fragmentation, and the strangeness suppression factor $\gamma_s$ 
are given by the following expressions, where $V\!M$ and $P\!S$ stand for
the number of vector mesons and pseudoscalar mesons, and $B\!R$ are
branching ratios:
\vspace*{-0.4cm}
\par\noindent
\begin{minipage}[t]{12cm}
\begin{eqnarray}
 \nonumber
 \begin{array}{lll}
 P_V &  = \ \frac{\displaystyle V\!M}{\displaystyle P\!S+V\!M}  
     &  = \ \frac{\displaystyle f(c \rightarrow D^{*+})}
{\displaystyle f(c\rightarrow D^+) +
     f(c\rightarrow D^{*+})\ BR(D^{*+}\rightarrow D^{\circ}\pi^+)} \\
 \end{array}
\end{eqnarray}
\end{minipage}
\vspace*{-0.1cm}
\par\noindent
\begin{minipage}[t]{12.6cm}
\begin{minipage}[t]{5.8cm}
\begin{eqnarray}
 \nonumber
 \begin{array}{ll}
  P^{\prime}_V & = \ \frac{\displaystyle 2\ f(c \rightarrow D^{*+})}
{\displaystyle f(c\rightarrow D^+) +f(c\rightarrow D^{\circ})}  \\
 \end{array}
\end{eqnarray}
\end{minipage}
\vspace*{-1.35cm}
\par\noindent
\hspace*{6.5cm}
\begin{minipage}[t]{5.5cm}
\begin{eqnarray}
\nonumber
 \begin{array}{ll}
   \gamma_s  & = \ \frac{\displaystyle 2\ f(c \rightarrow D_s^+)}
{\displaystyle f(c\rightarrow D^+) +f(c\rightarrow D^o)}\\
 \end{array}
\end{eqnarray}
\end{minipage}
\end{minipage}
\vspace*{-0.4cm}
\par\noindent
\begin{minipage}[t]{10cm}
\begin{eqnarray}
 \nonumber
 \begin{array}{ll}
R_{u/d}     & = \ \frac
{\displaystyle f(c\rightarrow D^{\circ}) -f(c\rightarrow D^{*+})\ BR(D^{*+}\rightarrow D^{\circ}\pi^+)}
{\displaystyle f(c\rightarrow D^+) +f(c\rightarrow D^{*+})\ BR(D^{*+}\rightarrow D^{\circ}\pi^+)}\\
 \end{array}
\end{eqnarray}   
\end{minipage}
\vspace*{0.2cm}
\par\noindent
In the ratio $P^{\prime}_V$ isospin invariance has been used, i.e. it is 
assumed that $f(c\rightarrow D^{*+}) = f(c\rightarrow D^{*{\circ}})$.
\par\noindent
The obtained numbers, $P_V = 0.693 \pm 0.045 \pm 0.004 \pm 0.009$,    
$R_{u/d} = 1.26 \pm 0.20 \pm 0.13 \pm 0.04$ and
$\gamma_s = 0.36 \pm 0.10 \pm 0.01 \pm 0.08$, agree well with the world
average values (calculated using \cite{Gladilin99})
of $0.601 \pm 0.032$, $1.00 \pm 0.09$    
and $0.26 \pm 0.07$, respectively. The ZEUS measurements\cite{ZEUSratios} of 
$\gamma_s = 0.27 \pm 0.05 \pm 0.07$ and  
$P^{\prime}_V  = 0.546 \pm 0.045 \pm 0.028$ also compare well with the 
world average and with the H1 measurement of 
$P^{\prime}_V = 0.613 \pm 0.061 \pm 0.033 \pm 0.008$, respectively.
\par\noindent
\begin{minipage}[t]{12.6cm}
\begin{minipage}[t]{5.5cm}
\epsfig{file=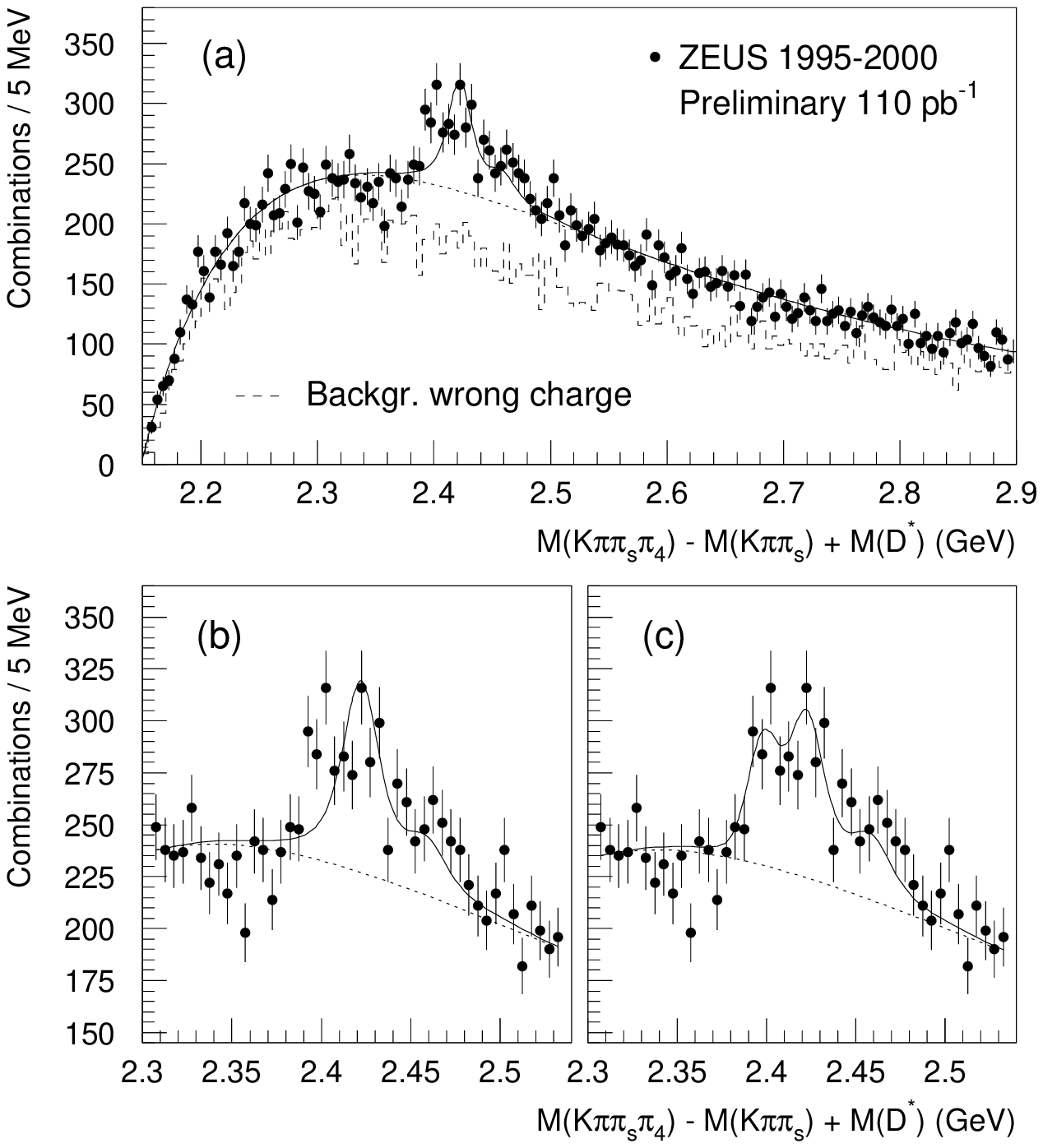,width=5.5cm,
      bbllx=1pt,bblly=1pt,bburx=380pt,bbury=425pt,clip= }
Figure 18: Mass distribution $\Delta\!M^{**} =
M(K\pi\pi_s\, \pi_4) - M(K\pi\pi_s) + M(D^{*\pm}(2010))$. The fits are 
described in the text.
\end{minipage}
\vspace*{0.3cm}
\par\noindent
\begin{minipage}[t]{5.5cm}
\epsfig{file=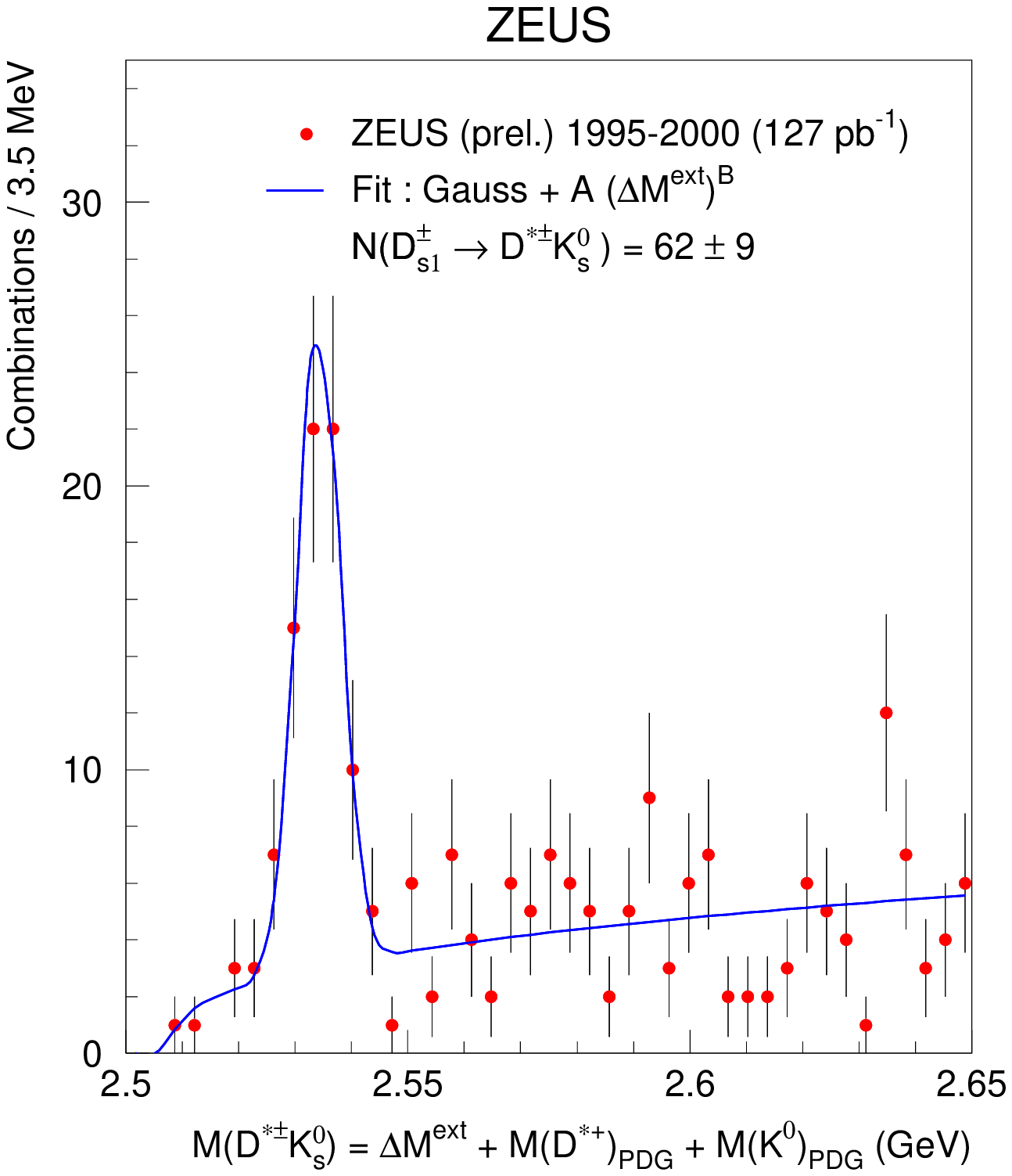,width=5.5cm,
      bbllx=1pt,bblly=1pt,bburx=375pt,bbury=445pt,clip= }
Figure 19: Mass distribution $\Delta\!M^{ext} =
M(K\pi\pi_s\, \pi_3\pi_4) - M(K\pi\pi_s) - M(\pi_3\pi_4)$ 
\end{minipage}
\vspace*{-16.3cm}
\par\noindent
\hspace*{6cm}
\begin{minipage}[t]{6.6cm}
The world average values of these fragmentation sensitive ratios 
are dominated by the LEP $e^+e^-$ results. Thus one may conclude that the 
hypothesis of charm fragmentation universality is well supported by the 
HERA $ep$ data. It is argued in \cite{Nason} that different processes may
be sensitive to different aspects of fragmentation, and that universality
may not hold.
\vspace*{0.1cm}
\par\noindent
{\bf Observation of \boldmath$P$-wave states: \ } ZEUS observe two neutral
states, using an ``extended $\Delta m$'' tagging method, 
in which still another pion with the correct charge was added to 
the $D^*$ candidate combination\cite{ZEUSneutralPwave}.
The corresponding mass distribution is shown in Fig.~18. A fit which includes
two Breit-Wigner shapes, folded with the resolution and including the expected
helicity spectra ($J^P=1^+$ and $2^+$), describes the data (Fig.~18b). The
two states agree in mass and width with the $L=1$ states previously seen in
$e^+e^-$ collisions, namely $D_1^{\circ}(2420)$
and $D_2^{*\circ}(2460)$\cite{PDG2000}. 
The fragmentation factors determined by ZEUS,
$f(c\rightarrow D_1^{\circ}) =
                    1.46 \pm 0.18 \ ^{+0.33}_{-0.27} \pm 0.06$ \% and
$f(c\rightarrow D_2^{*\circ}) =
                    2.00 \pm 0.58 \ ^{+1.40}_{-0.48} \pm 0.41$ \%, 
also agree with the previous measurements.
\vspace*{0.05cm}
\par\noindent
The mass spectrum in Fig.~18 is complicated by the indication of a third mass
state, not previously seen. Fig.~18c shows a fit with an added 
Breit-Wigner shape for a new hypothetical state at $\sim~2.4$ GeV. Is it 
an interference effect, a fluctuation, or indeed a new state?
\end{minipage}
\end{minipage}
\newpage 
\par\noindent
The strange-charmed $P$-wave state $D_{s1}^{\pm}(2536)$
is also observed by ZEUS\cite{ZEUSDsubS1}, in the 
``extended $\Delta m$'' distribution 
$\Delta\!M^{ext} =
M(K\pi\pi_s\, \pi_3\pi_4) - M(K\pi\pi_s) - M(\pi_3\pi_4)$, in which the 
two additional pions form a $K^{\circ}_S$. The statistics in Fig.~19, 
$62\pm 9$ events,
are not enough to establish the spin-parity, decay distributions are 
compatible both with $1^+, 1^-$ and $2^+$.
\vspace*{0.1cm}
\par\noindent
Nevertheless, both mass and width
and the fragmentation factor, 
$f(c\rightarrow D_{s1}^{\pm}) =
1.24 \pm 0.18 \ ^{+0.08}_{-0.06} \pm 0.14({\rm br.})$ \%, are in good 
agreement with the previous $e^+e^-$ observations\cite{PDG2000}. 
Not understood is the fact that  
$f(c\rightarrow D_{s1}^{\pm}) > \gamma_s \cdot  f(c\rightarrow  D_1^{\circ})$,
meaning that the strangeness suppression, which was determined using the 
$S$-wave states, seems not to be valid for this $P$-wave state.
\vspace*{0.1cm}
\par\noindent
\begin{minipage}[t]{12.6cm}
\vspace*{0.4cm}
\par\noindent
\begin{minipage}[t]{5cm}
\epsfig{file=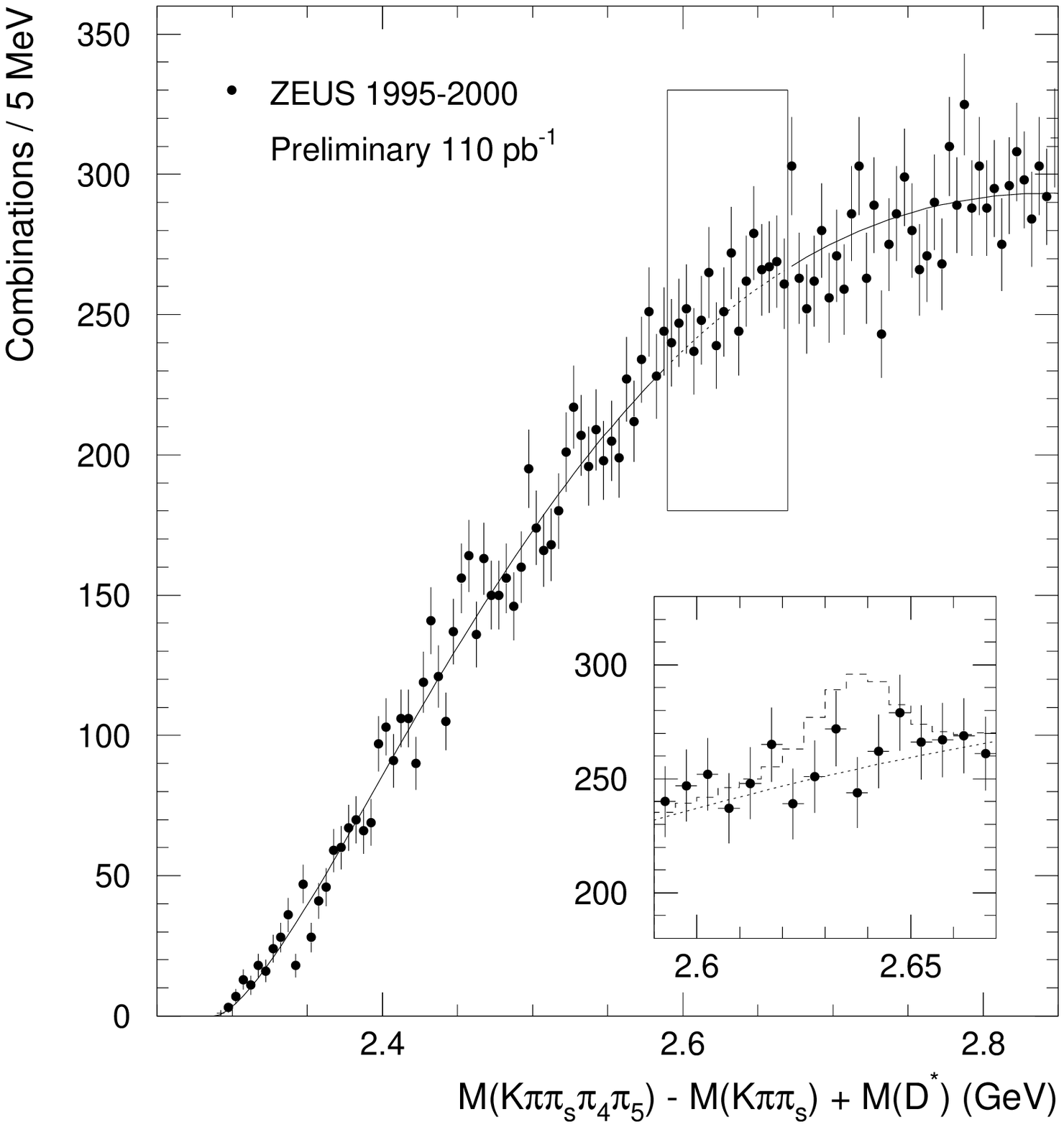,width=5cm,
      bbllx=1pt,bblly=1pt,bburx=470pt,bbury=510pt,clip= }
Figure 20: Mass distribution  $\Delta\!M^{*\prime} =
M(K\pi\pi_s\, \pi_4\pi_5) - M(K\pi\pi_s) + M(D^{*\pm})$. The insert shows the
signal expectation corresponding to the upper limit.
\end{minipage}
\vspace*{-9.2cm}
\par\noindent
\hspace*{5.5cm}
\begin{minipage}[t]{7cm}
Finally, ZEUS\cite{ZEUSneutralPwave} have also searched for the 
narrow state seen by 
DELPHI\cite{DELPHI_2637} at 2637 MeV and decaying into $D^{*\pm}\pi^+\pi^-$.
At this mass, radially excited $D^*$-mesons are expected\cite{GodfreyIsgur}.
The search method is again an ``extended $\Delta m$'' combination,   
$\Delta\!M^{*\prime} =
M(K\pi\pi_s\, \pi_4\pi_5) - M(K\pi\pi_s) + M(D^{*\pm})$, where two pions
$\pi^+\pi^-$ have been added to the $D^{*\pm}$ candidate combinations. The
mass spectrum in Fig.~20 shows no significant peak, and an upper limit is set,
\newline
$R_{D^{*\prime\pm} \rightarrow D^{*\pm}\pi^+\pi^- /D^{*\pm}} 
      < 2.3 \%   {\rm (95\% \ C.L.)}$, \newline
which contradicts the DELPHI 
observation. The upper limit can also be expressed as a limit on the 
corresponding fragmentation factor,\newline 
$f(c\rightarrow D^{*\prime\pm})~\cdot~BR(D^{*\prime +}\rightarrow~D^{*+}\pi^+\pi^-) \newline
<~0.7~\%\,{\rm (95\%~C.L.)}$. OPAL and CLEO \cite{OPALCLEO_2637} also could
not confirm the DELPHI observation, and OPAL gave the value 0.9~\%\ for 
this limit.
\end{minipage}
\end{minipage}
\vspace*{0.2cm}
\par\noindent
In conclusion, the Charm Spectroscopy is a very active field in the HERA
physics, and it will greatly profit from the HERA-II luminosity upgrade and 
from the recent H1 and ZEUS detector improvements, in particular
in the areas of triggering, tracking and vertex detection.
 
\section{Acknowledgments}
 
It is a pleasure to thank the organizers for the warm and joyful atmosphere 
in an exciting, interesting and very well prepared conference.
I also wish to thank my
colleagues in H1 and ZEUS, for providing the data and the results 
shown in this report and for all their help given to me. I am particularly 
indebted to L. Gladilin, D. Ozerov and Y. Yamazaki for critical remarks 
to the manuscript.


\begin{thebibliography}{99}
\renewcommand{\baselinestretch}{0.5}
{\footnotesize 
\bibitem{Crittenden} J.A. Crittenden, Springer Tracts in
          Modern Physics, Volume 140 (Springer, Berlin Heidelberg, 1997)
\bibitem{AbramowiczCaldwell} H. Abramowicz and A. Caldwell, 
            Rev.Mod.Phys.71 (1999) 1275
\bibitem{Brodsky94} S.J. Brodsky \etal, Phys. Rev. {\bf D50} (1994) 3134
\bibitem{Ryskin93} M.G. Ryskin, Z. Phys. {\bf C57} (1993) 89
\bibitem{H1ZEUS9596} M. Erdmann, Proceedings ICHEP1998,
Vancouver, Canada, Vol. 1, p. 217
\bibitem{H1Newrhodata} H1 Collab., presented by X. Janssen in DIS2002,
                 Krak\'{o}w, Poland, 
http://www-h1.desy.de/psfiles/confpap/DIS2002/H1prelim-02-015.ps
\bibitem{DL84} A. Donnachie and P.V. Landshoff, 
                       Nucl.Phys. {\bf B231} (1984) 189
\bibitem{H1ZEUSJPsidata} ZEUS Collab., S. Chekanov \etal, 
                         Eur.Phys.J. {\bf C24} (2002) 345; \newline
                       ZEUS Collab., presented by A. Levy in DIS2002, 
               Krak\'{o}w, Poland; \newline
   H1 Collab., C. Adloff \etal,  Eur.Phys.J. {\bf C10} (1999) 373; \newline 
   H1 Collab., C. Adloff \etal, Phys.Lett {\bf B483} (2000) 23                
\bibitem{FKS98} L. Frankfurt, W. Koepf and M. Strikman, Phys.~Rev.~{\bf D57}
 (1998) 512
\bibitem{MRT99} A.D. Martin, M.G. Ryskin and T. Teubner, Phys. Rev. {\bf D26}
(1999) 14022 
\bibitem{CTEQ45M} CTEQ Collab. H.L. Lai \etal, 
                Eur.Phys.J. {\bf C12} (2000) 375; \newline
                  CTEQ Collab. H.L. Lai \etal, 
                  Phys.Rev.. {\bf D55} (1997) 1280
\bibitem{Browndis2001} H1 Collab., presented by D. Brown in DIS2001, Bologna,
Italy, 
http://www-zeus.desy.de/conferences/01/2001\_DIS\_brown\_writeup.ps.gz
\bibitem{Forshaw96} J. Bartels \etal, Phys.Lett. {\bf B375} (1996) 301
\bibitem{ZEUSLT} ZEUS Collab., presented by A. Kreisel in DIS2001, Bologna,
Italy, 
http://www-zeus.desy.de/conferences/01/2001\_DIS\_kreisel.ps.gz 
\bibitem{FKS96} L. Frankfurt, W. Koepf and M. Strikman, Phys.~Rev.~{\bf D54}
 (1996) 3194
\bibitem{newhightdata} ZEUS Collab., S. Chekanov \etal, Preprint DESY-02-072,
                         hep-ex/0205081
\bibitem{Ivanov2000} D.Yu. Ivanov \etal, Phys.Lett. {\bf B478} (2000) 101;
     Err. Phys.Lett. {\bf B498} (2001) 295 
\bibitem{Forshawnew} J.R. Forshaw and G. Poludniowski, Preprint hep-ph/0107068
\bibitem{BFKL} E.A. Kuraev, L.N. Lipatov and V.S. Fadin, 
                 Sov.Phys.JETP {\bf 45} (1977) 199; \newline
      Ya.Ya. Balitski\u{i} and L.N. Lipatov, 
Sov.J.Nucl.Phys. {\bf 28} (1978) 822
\bibitem{Forshaw95}
           J.R. Forshaw and M. Ryskin, Z.Phys. {\bf C68} (1995) 137 
\bibitem{RCKNZ} I. Royen and J. Cudell, Nucl.Phys. {\bf B5454} (1999) 505;
        \newline
                I. Royen, Phys.Lett. {\bf B513} (2001) 337;
         \newline
E.A. Kuraev, N.N. Nikolaev and B.G. Zakharov, JETP Lett. {\bf 68} (1998) 696
\bibitem{IvanovKirschner} D.Yu.Ivanov and R. Kirschner, Phys. Rev. {\bf D58}
                              (1998) 114026
\bibitem{SWolf} K. Schilling and G. Wolf, Nucl. Phys. {\bf B61} (1973) 381;
\newline
for a detailed discussion, see also [1] 
\bibitem{H1elprodrho} ZEUS Collab., J. Breitweg \etal, 
                      Eur.Phys.J. {\bf C12} (2000) 393; \newline
                      H1 Collab., C. Adloff \etal, 
                      Eur.Phys.J. {\bf C13} (2000) 371
\bibitem{ZEUSelprodphi} ZEUS Collab.,
                      contr.~paper~793~to~ICHEP~1998, Vancouver, Canada
\bibitem{H1rhohight} H1 Collab., 
                  C. Adloff \etal, Phys.Lett. {\bf B539} (2002) 25
\bibitem{MRT} A.D. Martin, M.G. Ryskin and T. Teubner, Phys. Lett. {\bf B454}
(1999) 339
\bibitem{LUK73}  L. \L ukaszuk and B. Nicolescu,
                Lett.Nuov.Cim. {\bf 8} (1973) 405; \newline
                K. Kang and B. Nicolescu, Phys.Rev. {\bf D11} (1975)
                2461; \newline
                G. Bia\l kowski, K. Kang and B. Nicolescu,
                Lett.Nuov.Cim. {\bf 13} (1975) 401; \newline
                D. Joynson \etal, Nuov.Cim. {\bf 30A} (1975) 345
 
\bibitem{INTERF} S.J. Brodsky, J. Rathsman and C. Merino, Phys.Lett.
                 {\bf B461} (1999) 114; \newline
                 A. Ahmedov \etal, Eur.Phys.J. {\bf C11} (1999) 703;
         \newline
                 I.P. Ivanov, N.N. Nikolaev and I.F. Ginzburg, hep-ph/0110181;
         \newline
                 I.F. Ginzburg, I.P. Ivanov and N.N. Nikolaev, hep-ph/0207345;
         \newline
                 Ph. H\"agler \etal, Phys.Lett. {\bf B535} (2002) 117; 
             Err. Phys.Lett. {\bf B540} (2002) 324;
         \newline
                 Ph. H\"agler \etal, hep-ph/0207224, 
                             to be published in Eur.Phys.J.
\bibitem{BAR2001} J. Bartels \etal, Eur.Phys.J. {\bf C20} (2001) 323  
\bibitem{ERB99}  E.R. Berger \etal, Eur.Phys.J. {\bf C9} (1999) 491;
                 \newline 
                 E.R. Berger \etal, Eur.Phys.J. {\bf C14} (2000) 673;
                 \newline
                 H.G. Dosch, private communication, Heidelberg (2001) 
\bibitem{DoschNachtmann}  H.G. Dosch and Yu. A. Simonov,
                    Phys.Lett. {\bf B205} (1988) 339; \newline
                 O. Nachtmann, Ann.Phys. {\bf 209} (1991) 436
\bibitem{DoschDonnachie2002} 
              A. Donnachie and H.G. Dosch, Phys.Rev. {\bf D65} (2002) 014019
\bibitem{BUD01} H1 Collab., contr.~paper~795 to~IECHEP~2001 
                Budapest, Hungary 
\bibitem{H1Odderonpi0} H1 Collab., C. Adloff \etal, 
                DESY preprint 02-087, hep-ex/0206073, to be published in 
                Phys.Lett. {\bf B} 
\bibitem{Kaidalov99} 
      A.B. Kaidalov and Yu.A. Simonov, Phys.Lett. B477 (2000) 163
\bibitem{Busseyhq} P.J. Bussey, Int.J.Mod.Phys. {\bf A17} (2002) 1065
\bibitem{NewH1charmdata} H1 Collab., presented by J. Wagner in 
          DIS2002, Krak\'{o}w, Poland, \newline 
http://www-h1.desy.de/psfiles/confpap/DIS2002/H1prelim-02-076.ps 
\bibitem{AROMA} G. Ingelman, J. Rathsman and G.A. Schuler, 
                 Comput.Phys.Commun. {\bf 101} (1997) 135
\bibitem{Gladilin99} L. Gladilin, ``Charm Hadron Production Fractions'', 
                  hep-ex/9912064
\bibitem{ZEUSratios} ZEUS Collab., J. Breitweg \etal, Phys.Lett. {\bf B481}
                        (2000) 213; \newline
                ZEUS Collab., contr.~paper~501 to~IECHEP~2001,
                Budapest, Hungary
\bibitem{Nason} S. Frixione \etal, J.Phys. {\bf G27} (2001) 27; \newline
                P. Nason \etal, in LHC workshop on Standard Model Physics,
                hep-ph/0003142
\bibitem{ZEUSneutralPwave} ZEUS Collab., contr.~paper~448 (Abstr. 854) 
             to~ICHEP~2000, Osaka, Japan
\bibitem{PDG2000} Particle Data Group, D.E. Groom \etal, 
                  Eur.Phys.J. {\bf C15} (2000) 1
\bibitem{ZEUSDsubS1}  ZEUS Collab., contr.~paper~497 to~IECHEP~2001,
                Budapest, Hungary
\bibitem{DELPHI_2637} DELPHI Collab., P. Abreu \etal, Phys.Lett. {\bf B426}
                    (1998) 231
\bibitem{GodfreyIsgur} S. Godfrey and N. Isgur, Phys.Rev. {\bf D32} (1985) 189;
           \newline
                D. Ebert \etal, Phys.Rev. {\bf D32} (1998) 5663
\bibitem{OPALCLEO_2637} OPAL Collab, G. Abbiendi \etal, 
                 Eur.Phys.J. {\bf C20} (2001) 445; \newline
         CLEO Collab., presented by J.L. Rodriguez in HQ98, hep-ex/9901008   
}
\end{thebibliography}
\end{document}